\pdfoutput=1  

\documentclass[journal]{IEEEtran}
\usepackage{ifpdf}
\ifCLASSINFOpdf
  \usepackage[pdftex]{graphicx}
  \DeclareGraphicsExtensions{.pdf,.jpeg,.png}
\else
  \usepackage[dvips]{graphicx}
  \DeclareGraphicsExtensions{.eps}
\fi
\graphicspath{{./figures/}}
\usepackage{amsmath}
\usepackage{units} 
\newcommand\numberthis{\addtocounter{equation}{1}\tag{\theequation}} 

\usepackage{array}

\usepackage{multirow}
\usepackage{tabularx}
\usepackage{booktabs}
\newcolumntype{P}[1]{>{\centering\arraybackslash}p{#1}} 
\newcolumntype{L}[1]{>{\raggedright\let\newline\\\arraybackslash}p{#1}}
\newcolumntype{C}[1]{>{\centering\let\newline\\\arraybackslash}m{#1}}
\newcolumntype{R}[1]{>{\raggedleft\let\newline\\\arraybackslash}m{#1}}
\usepackage[hidelinks]{hyperref} 
\usepackage{dblfloatfix}

\usepackage{color}
\usepackage[nolist]{acronym} 

\hypersetup{pdfprintscaling=None}

\usepackage{flushend} 

\usepackage[symbols,nogroupskip,sort=use]{glossaries-extra}  
\makenoidxglossaries
\glsxtrnewsymbol[description={Sampled knee acoustic emission signal of the $i^{\text{th}}$ knee.}]{iKneeSig}{\ensuremath{s_i(n)}}

\glsxtrnewsymbol[description={Segment $j$ of the knee acoustic emission signal of the $i^{\text{th}}$ knee.}]{iKneeSigjSegm}{\ensuremath{s_{i,j}(n)}}

\glsxtrnewsymbol[description={Discrete time index.}]{DTindex}{\ensuremath{n}}

\glsxtrnewsymbol[description={Frequency index.}]{freqindex}{\ensuremath{k}}

\glsxtrnewsymbol[description={Sampling frequency.}]{Fs}{\ensuremath{F_s}}

\glsxtrnewsymbol[description={Imaginary unit.}]{imagI}{\ensuremath{\rm{i}}}

\glsxtrnewsymbol[description={The ratio of the circumference of a circle to its diameter (mathematical constant).}]{pi}{\ensuremath{\pi}}

\glsxtrnewsymbol[description={Window in seconds within which the sounds related to abnormalities appear in a knee acoustic emission signal.}]{assumpWindow}{\ensuremath{\tau_s}}

\glsxtrnewsymbol[description={Number of filters in a filter-bank.}]{numFilters}{\ensuremath{N_B}}

\glsxtrnewsymbol[description={Filter-bank matrix of triangular filters linearly spaced in frequency (in columns).}]{linFiltbankMat}{\ensuremath{\boldsymbol{U}_L}}

\glsxtrnewsymbol[description={Filter-bank matrix of triangular filters equally spaced along the mel-frequency axis (in columns).}]{melFiltbankMat}{\ensuremath{\boldsymbol{U}_M}}

\glsxtrnewsymbol[description={Number of frames in a signal.}]{numFrames}{\ensuremath{N_f}}

\glsxtrnewsymbol[description={Size in ms of a general frame of a signal.}]{genFrameSizems}{\ensuremath{l}}

\glsxtrnewsymbol[description={Size in samples of a general frame of a signal.}]{genFrameSizeSamps}{\ensuremath{l_n}}

\glsxtrnewsymbol[description={Matrix of frames from a sampled knee acoustic emission signal.}]{enframedSigMat}{\ensuremath{\boldsymbol S}}

\glsxtrnewsymbol[description={Discrete fourier transform of $\boldsymbol S$.}]{enframedSigMatDft}{\ensuremath{\boldsymbol{\Psi_F}}}

\glsxtrnewsymbol[description={Discrete fourier transform of $\boldsymbol S$ multiplied with a filter-bank matrix with triangular band-pass filters linearly spaced in frequency.}]{LinDftMat}{\ensuremath{\boldsymbol{\Psi_E}}}

\glsxtrnewsymbol[description={Discrete fourier transform of $\boldsymbol S$ multiplied with a filter-bank matrix with triangular band-pass filters equally spaced on the mel-frequency axis.}]{MelDftMat}{\ensuremath{\boldsymbol{\Psi_D}}}

\glsxtrnewsymbol[description={Hanning window diagonal matrix.}]{hanningDiagMat}{\ensuremath{\boldsymbol H}}

\glsxtrnewsymbol[description={Vandermonde matrix for the roots of unity.}]{dftMatCoefs}{\ensuremath{\boldsymbol W}}

\glsxtrnewsymbol[description={Operator that acts along the columns of a matrix and extracts 11 statistical parameters.}]{statExtractOper}{\ensuremath{\mathcal{D}}}

\glsxtrnewsymbol[description={Discrete Cosine Transform operator.}]{DCToper}{\ensuremath{\mathcal{C}}}

\glsxtrnewsymbol[description={MFCC feature vector of \boldsymbol{\Psi_D}.}]{mfccPsiD}{\ensuremath{\boldsymbol{C_M}}}

\glsxtrnewsymbol[description={LFCC feature vector of \boldsymbol{\Psi_E}.}]{LfccPsiE}{\ensuremath{\boldsymbol{C_L}}}

\glsxtrnewsymbol[description={Statistical feature vector of $\boldsymbol{\Psi_D}$.}]{MelFreqSTFTfeatVec}{\ensuremath{\boldsymbol{\phi_D}}}  
\glsxtrnewsymbol[description={Matrix comprised of a collection of $\boldsymbol{\phi_D}$ vectors for a number of knee acoustic emission signal segments.}]{MelFreqSTFTfeatSet}{\ensuremath{\boldsymbol{\Phi_D}}}

\glsxtrnewsymbol[description={Statistical feature vector of $\boldsymbol{\Psi_E}$.}]{LinFreqSTFTfeatVec}{\ensuremath{\boldsymbol{\phi_E}}}  
\glsxtrnewsymbol[description={Matrix comprised of a collection of $\boldsymbol{\phi_E}$ vectors for a number of knee acoustic emission signal segments.}]{LinFreqSTFTfeatSet}{\ensuremath{\boldsymbol{\Phi_E}}}

\glsxtrnewsymbol[description={Statistical feature vector of $\boldsymbol{\Psi_F}$.}]{STFTfeatVec}{\ensuremath{\boldsymbol{\phi_F}}}  
\glsxtrnewsymbol[description={Matrix comprised of a collection of $\boldsymbol{\phi_F}$ vectors for a number of knee acoustic emission signal segments.}]{STFTfeatSet}{\ensuremath{\boldsymbol{\Phi_F}}}

\glsxtrnewsymbol[description={Statistical feature vector of $\boldsymbol{C_M}$.}]{mfccfeatVec}{\ensuremath{\boldsymbol{\phi_M}}}  
\glsxtrnewsymbol[description={Matrix comprised of a collection of $\boldsymbol{\phi_M}$ vectors for a number of knee acoustic emission signal segments.}]{mfccfeatSet}{\ensuremath{\boldsymbol{\Phi_M}}}

\glsxtrnewsymbol[description={Statistical feature vector of $\boldsymbol{C_L}$.}]{LfccfeatVec}{\ensuremath{\boldsymbol{\phi_L}}}  
\glsxtrnewsymbol[description={Matrix comprised of a collection of $\boldsymbol{\phi_L}$ vectors for a number of knee acoustic emission signal segments.}]{LfccfeatSet}{\ensuremath{\boldsymbol{\Phi_L}}}

\glsxtrnewsymbol[description={Error rate.}]{errRate}{\ensuremath{\text{E}_{r}}}

\glsxtrnewsymbol[description={Harmonic mean of precision and recall.}]{f1score}{\ensuremath{\text{F}_1}}

\glsxtrnewsymbol[description={Variation of $\text{F}_1$ which weighs recall lower than precision.}]{f05score}{\ensuremath{\text{F}_{0.5}}}

\glsxtrnewsymbol[description={Threshold value for the error rate.}]{errRateThres}{\ensuremath{\theta_{er}}}

\glsxtrnewsymbol[description={Threshold value for the $\text{F}_{0.5}$ score.}]{f05scoreThres}{\ensuremath{\theta_{0.5}}}

\glsxtrnewsymbol[description={Threshold value for the Matthew's Correlation Coefficient.}]{mccThres}{\ensuremath{\theta_{mcc}}}

\glsxtrnewsymbol[description={Average of $\text{E}_{r}$, $\text{F}_{0.5}$ and MCC (a classification metric) .}]{s_score}{\ensuremath{\text{S}_\text{c}}}

\makeatletter
\AtBeginDocument{%
	\renewcommand*{\AC@hyperlink}[2]{%
		\begingroup
		\hypersetup{hidelinks}%
		\hyperlink{#1}{#2}%
		\endgroup
	}%
}
\makeatother

\begin{document}
\bstctlcite{IEEEexample:BSTcontrol} 
%
\title{Time-Frequency Analysis and Parameterisation of Knee Sounds for Non-invasive Detection of Osteoarthritis}  
%
%
%

\author{Costas~Yiallourides
        and~Patrick~A~Naylor,~\IEEEmembership{Senior Member,~IEEE}
\thanks{This work has been submitted to the IEEE for possible publication. Copyright \textcopyright 2020 IEEE. Personal use of this material is permitted. Permission from IEEE must be obtained for all other uses, in any current or future media, including reprinting/republishing this material for advertising or promotional purposes, creating new collective works, for resale or redistribution to servers or lists, or reuse of any copyrighted component of this work in other works.}
\thanks{C. Yiallourides and P. A. Naylor are with the Department
	of Electrical and Electronic Engineering, Imperial College London,
	UK. \href{mailto:costas.yiallourides08@imperial.ac.uk}{costas.yiallourides08@imperial.ac.uk}, \href{mailto:p.naylor@imperial.ac.uk}{p.naylor@imperial.ac.uk}}
}

%
%

%
\markboth{A PREPRINT}%
{Shell \MakeLowercase{\textit{et al.}}: Bare Demo of IEEEtran.cls for IEEE Journals}



\begin{acronym}
\acro{3GPP}{3rd Generation Partnership Project}
\acro{A-SNR}{A-weighted \ac{SNR}}
\acro{AAC}{Advanced Audio Coding\acroextra{. A lossy codec used for digital audio.}}
\acro{AASP}{Audio and Acoustic Signal Processing}
\acro{ACAWD}{Archivable Core Actual-Word Database}
\acro{ACB}{Adaptive Codebook}
\acro{ACC}{Accuracy}
\acro{ACE}{Acoustic Characterization of Environments\acroextra{. An IEEE challenge run by the SAP group at Imperial College}}
\acro{ACELP}{Algebraic Code-Excited Linear Prediction}
\acro{ACR}{Absolute Category Rating}
\acro{AD}{Audio Diarization}
\acro{ADC}{Analogue-to-Digital Converter}
\acro{ADPCM}{Adaptive Differential Pulse Code Modulation}
\acro{AE}{Almost Everywhere}
\acro{AE1}[AE]{Acoustic Emission}
\acro{AES}{Audio Engineering Society}
\acro{AGC}{Automatic Gain Control}
\acro{AH}{Amplitude Histogram}
\acro{AI}{Articulation Index}
\acro{AI2}[AI]{Artificial Intelligence}
\acro{AI3}[AI]{Audio Inpainting}
\acro{AIC}{Akaike Information Criterion}
\acro{AIFF}{Audio Interchange File Format}
\acro{AIR}{Acoustic Impulse Response}
\acro{AIRD}{Aachen Impulse Response Database}
\acro{ALC}{Automatic Level Control}
\acro{ALCons}{Articulation Loss of Consonants}
\acro{AM}{Amplitude Modulation}
\acro{AMDF}{Average Magnitude Difference Function\acroextra{. A function with similar properties to the cross- or autocorrelation but that requires no multiplication to evaluate.}}
\acro{AMR}{Adaptive Multi-Rate}
\acro{AMR-NB}{Adaptive Multi-Rate Narrow Band}
\acro{AMR-WB}{Adaptive Multi-Rate Wide Band}
\acro{ANC}{Adaptive Noise Canceller}
\acro{ANS}{Autocorrelation-Based Noise Subtraction}
\acro{ANSI}{American National Standards Institute}
\acro{APLAWD}{Archivable Priority List Actual-Word Database}
\acro{AR}{Autoregressive}
\acro{ARD}{Arbeitsgemeinschaft der \"{o}ffentlich-rechtlichen Rundfunkanstalten der Bundesrepublik Deutschland\acroextra{. `Consortium (``Working group'') of the public-law broadcasting institutions of the Federal Republic of Germany'}}
\acro{ARMA}{Autoregressive Moving Average}
\acro{AS}{Audio Segmentation}
\acro{AS2}[AS]{Almost Surely}
\acro{ASA}{Acoustic Scene Analysis}
\acro{ASIO}{Audio Stream Input/Output\acroextra{. A computer soundcard protocol with low latency developed by Steinberg.}}
\acro{ASK}{Amplitude Shift Keying}
\acro{ASLP}{Audio, Speech, and Language Processing}
\acro{ASR}{Automatic Speech Recognition}
\acro{ASS}{Approximate Spectrum Substitution}
\acro{ATLM}{Acoustic Tokenization and Language Modelling}
\acro{ATR}{Advanced Telecommunications Research Institute International\acroextra{, Kyoto, Japan}}
\acro{AUC}{area under the curve}
\acro{AWGN}{Additive White Gaussian Noise}
\acro{AWS}{Approximate Waveform Substitution}
\acro{BASIE}{Bayesian Adaptive Speech Intelligibility Estimation}
\acro{BCE}{Blind Channel Estimation}
\acro{BCL}{Bekesy Comfortable Loudness}
\acro{BCR}{Block-Coordinate Relaxation}
\acro{BEM}{Boundary Element Method}
\acro{BER}{Bit Error Rate}
\acro{BIBO}{Bounded-Input Bounded-Output}
\acro{BIC}{Bayesian Information Criterion}
\acro{Bk}{Berksons\acroextra{. A unit for measuring intelligibility.}}
\acro{BP}{Basis Pursuit}
\acro{BPCC}{Basis Pursuit with Clipping Constraints}
\acro{BPDN}{Basis Pursuit Denoising}
\acro{BPM}{Beats Per Minute}
\acro{BPSK}{Binary Phase Shift Keying}
\acro{BR}{Barrodale and Roberts' (algorithm)}
\acro{BRI}{Basic Rate Index}
\acro{BSD}{Bark Spectral Distortion}
\acro{BSI}{Blind System Identification}
\acro{BSS}{Blind Source Separation}
\acro{BW}{Bandwidth}
\acro{BZ}{Back-to-Zero}
\acro{CART}{Classification and Regression Tree}
\acro{CASA}{Computational Auditory Scene Analysis}
\acro{CBR}{Constant Bit Rate}
\acro{CCC}{Cross-Correlation Coefficient}
\acro{CCCC}{DARPA CSR Corpus Coordinating Committee}
\acro{CCD}{Charge-Coupled Device}
\acro{CCI}{Call Clarity Index}
\acro{CCITT}{Consultative Committee for International Telephony and Telegraphy}
\acro{CCM}{Contralateral Competing Message}
\acro{CCR}{Comparison Category Rating}
\acro{CDB}{Constant Directivity Beamformer}
\acro{CDF}{Cumulative Distribution Function}
\acro{CDMA}{Code Division Multiple Access}
\acro{CELP}{Code-excited Linear Prediction}
\acro{CIT}{Constrained Initial Taps}
\acro{CL}{Clipping Level}
\acro{CLEAR}{Centre for Law Enforcement Audio Research}
\acro{CLID}{Cluster Identification Test}
\acro{CLT}{Central Limit Theorem}
\acro{CMA}{Constant Modulus Algorithm}
\acro{CMB}{Cosmic Microwave Background}
\acro{CNC}{Consonant-Nucleus-Consonant}
\acro{CNG}{Comfort Noise Generation}
\acro{CODEC}{Coder-Decoder}
\acro{CPHD}{Cardinalized \ac{PHD}}
\acro{CRACD}{Codec-robust Automatic Clipping Detector}
\acro{CS}{Channel Shortening}
\acro{CS2}[CS]{Compressive Sensing}
\acro{CSP}{Communications and Signal Processing}
\acro{CSR-WSJ}{Continuous Speech Recognition Wall Street Journal Phase 1\acroextra{ database}}
\acro{CST}{Connected Speech Test\acroextra{ speech corpus}}
\acro{CT}{Conversation Test}
\acro{CTTN}{Comparative Tolerance to Noise}
\acro{CV}{Coefficient of Variation}
\acrodefplural{CV}{Coefficients of Variation}
\acro{CVC}{Consonant-Vowel-Consonant}
\acro{CW}{Continuous Wave}
\acro{CWM}{Centre-Weighted Median}
\acro{CWMY}{Centre-Weighted Myriad}
\acro{CWT}{Continuous Wavelet Transform}
\acro{DAC}{Digital-to-Analogue Converter}
\acro{DAM}{Diagnostic Acceptability Measure}
\acro{DAQ}{Data Acquisition}
\acro{DARPA}{Defense Advanced Research Projects Agency\acroextra{ of the United States Dept. of Defense}}
\acro{DARPA-RMD}{DARPA 1000-Word Resource Management Database\acroextra{ for Continuous Speech Recognition}}
\acro{DAW}{Digital Audio Workstation}
\acro{dB}{Decibel}
\acro{dBFS}{\ac{dB} Full Scale}
\acro{DC}{Direct Current}
\acro{DCME}{Digital Circuit Multiplexing Equipment}
\acro{DCR}{Degradation Category Rating}
\acro{DCT}{Discrete Cosine Transform}
\acro{DDR3}{Double Data Rate Type Three}
\acro{DECT}{Digital European Cordless Telecommunication}
\acro{DeLILAH}{Detection of Clipping using Least Squares Residuals and Iterated Logarithm Amplitude Histogram}
\acro{DET}{Detection Error Trade-off}
\acro{Dev}{Development\acroextra{ dataset of the \ac{ACE} Challenge}}
\acro{DFT}{Discrete Fourier Transform}
\acro{DI}{Directivity Index}
\acro{DirectX}{\acroextra{A programming interface developed by Microsoft for handling tasks related to multimedia.}}
\acro{DMA}{Differential Microphone Array}
\acro{DMT}{Discrete Multi-Tone}
\acro{DMV}{Dynamically Managed Voice\acroextra{ system}}
\acro{DNN}{Deep Neural Network}
\acro{DNR}{Dynamic Noise Reduction}
\acro{DoA}{Direction-of-Arrival}
\acrodefplural{DoA}[DoAs]{Directions-of-Arrival}
\acro{DOA}{Direction-of-Arrival}
\acro{DP}{Dynamic Programming}
\acro{DPCM}{Differential Pulse Code Modulation}
\acro{DPD}{Direct-Path Dominance}
\acro{DR}{Douglas-Rachford}
\acro{DR2}{Dynamic Range}
\acro{DRM}{Diagnostic Rhyme Test}
\acro{DRR}{Direct-to-Reverberant Ratio}
\acro{DRT}{Diagnostic Rhyme Test}
\acro{DSB}{Delay-and-Sum Beamformer}
\acro{DSOBM}{Deterministic STOI-optimal Binary Mask}
\acro{DSP}{Digital Signal Processing}
\acro{DSPS}{Double Sides Periodic Substitution}
\acro{DSR}{Distributed Speech Recognition}
\acro{DSWS}{Double Sides Waveform Substitution}
\acro{DTMF}{Dual Tone Multi-Frequency}
\acro{DTX}{Discontinued Transmission}
\acro{DWT}{Discrete Wavelet Transform}
\acro{EARS}{Embodied Audition for RobotS}
\acro{EBF}{Eigen-beamformer}
\acro{EBU}{European Broadcasting Union}
\acro{EC}{Echo Canceller}
\acro{EDC}{Energy Decay Curve}
\acro{EDR}{Energy Decay Relief}
\acro{EEG}{Electroencephalography}
\acro{EER}{Equal Error Rate}
\acro{EFICA}{Efficient Fast Independent Component Analysis}
\acro{EIR}{Equalized Impulse Response}
\acro{EKF}{Extended Kalman Filter}
\acro{EL}{Echo Loss}
\acro{ELF}{Extremely Low Frequency}
\acro{ELRA}{European Languages Research Association}
\acro{EM}{Estimation-Maximization\acroextra{. An iterative technique to solve certain optimization problems.}}
\acro{EPSRC}{Engineering and Physical Sciences Research Council}
\acro{EQ}{Equalisation}
\acro{ERB}{Equivalent Rectangular Bandwidth}
\acro{ERP}{Ear Reference Point (cf. ITU-T Rec. P.64 1999)}
\acro{ESA}{Early Stage Assessment}
\acro{ESPRIT}{Estimation of Signal Parameters via Rotational Invariance Techniques}
\acro{ETAN}{Equivalent Tolerance to Additional Noise} 
\acro{ETSI}{European Telecommunications Standards Institute}
\acro{EURASIP}{European Association for Signal Processing}
\acro{EUSIPCO}{European Signal Processing Conference}
\acro{Eval}{Evaluation\acroextra{ dataset of the \ac{ACE} Challenge}}
\acro{F1}{F1 Score}
\acro{FastSLAM}{FActored Solution To \ac{SLAM}} 
\acro{FB}{Forward-Backward}
\acro{FBF}{Fixed Beamformer}
\acro{FBSS}{Forward-Backward Spatial Smoothing}
\acro{FCC}{Federal Communications Commission}
\acro{FDM}{Frequency Division Multiplexing}
\acro{FDR}{False Discovery Rate}
\acro{FDR2}[FDR]{Free-Decay Region}
\acro{FEC}{Forward Error Correction}
\acro{FEM}{Finite Element Method}
\acro{FFT}{Fast Fourier Transform}
\acro{FIFO}{First-In First-Out}
\acro{FIR}{Finite Impulse Response\acroextra{. A filter whose output is a weighted sum of past input values and whose system function contains only zeros and no poles.}}
\acro{FISM}{Fast Image Source Method}
\acro{FISST}{Finite Set STatistics}
\acro{FLOM}{Fractional Lower-Order Moments}
\acro{FLOS}{Fractional Lower-Order Statistics}
\acro{FM}{Frequency Modulation}
\acro{FN}{False Negative}
\acro{FN2}{Nth Formant}
\acro{FNR}{False Negative Rate}
\acro{FORTRAN}{The IBM Mathematical Formula Translating System}
\acro{FoV}{Field of View}
\acro{FP}{False Positive}
\acro{FPR}{False Positive Rate}
\acro{FPS}{Frames Per Second}
\acro{FRI}{Finite Rate of Innovation}
\acro{FSB}{Filter-and-Sum Beamformer}
\acro{FSK}{Frequency Shift Keying}
\acro{FT}{Flat-Top}
\acro{FWER}{Familywise Error Rate}
\acro{G.711}{\ac{PCM} of Voice Frequencies}
\acro{GARCH}{Generalized Auto-regressive Conditional Heteroscedasticity}
\acro{GBW}{Gain Bandwidth Product}
\acro{GCC}{Generalized Cross-Correlation}
\acro{GGD}{Generalized Gaussian Distribution}
\acro{GGD2}[G$\Gamma$D]{Generalised Gamma Distribution}
\acro{GM}{Gaussian Mixture}
\acro{GMCA}{Generalized Morphological Component Analysis}
\acro{GMM}{Gaussian Mixture Model\acroextra{. An approximation to an arbitrary probability density function that consists of a weighted sum of Gaussian distributions}}
\acro{GM-PHD}{Gaussian Mixture \ac{PHD}}
\acro{GPRS}{General Packet Radio Services}
\acro{GSC}{Generalized Sidelobe Canceller}
\acro{GSM}{Global System For Mobile Communications}
\acro{GSM-EFR}{\ac{GSM} Enhanced Full Rate Codec}
\acro{GSM-FR}{\ac{GSM} Full Rate Codec}
\acro{GSM-HR}{\ac{GSM} Half Rate Codec}
\acro{GUI}{Graphical User Interface}
\acro{HAAC}{High Amplitude Audio Capture}
\acro{HATS}{Head and Torso Simulator}
\acro{HERB}{Harmonicity-based dEReverBeration}
\acro{HFT}{Hands-Free Terminal}
\acro{HINT}{Hearing-in-Noise Test}
\acro{HLT}{Human Language Technology}
\acro{HMM}{Hidden Markov Model}
\acro{HOS}{Higher-Order Statistics}
\acro{HPF}{High-Pass Filter}
\acro{HR}{Half Rate (\ac{GSM} Codec)}
\acro{HRI}{Human-Robot Interaction}
\acro{HRTF}{Head-related Transfer Function}
\acro{HSD}{Hybrid Steepest Descent}
\acro{HTK}{Hidden Markov Model Tool Kit}
\acro{IBM}{Ideal Binary Mask}
\acro{IC}{Interference Canceller}
\acro{ICA}{Independent Component Analysis}
\acro{ICASSP}{IEEE International Conference on Acoustics, Speech and Signal Processing}
\acro{ID}{Identifier}
\acro{iDEN}{Integrated Digital Enhanced Network}
\acro{IEC}{International Electrotechnical Commission}
\acro{IEEE}{Institute of Electrical and Electronics Engineers}
\acro{IET}{Institute of Engineering and Technology}
\acro{IETF}{Internet Engineering Task Force}
\acro{IFFT}{Inverse Fast Fourier Transform}
\acro{IHC}{Inner Hair Cell}
\acro{IHT}{Iterative Hard Thresholding}
\acro{iid}[i.i.d.]{Independent and Identically Distributed}
\acro{IID}{Independent and Identically Distributed}
\acro{IIR}{Infinite Impulse Response\acroextra{. A filter whose output is a weighted sum of both past input and past output values and whose system function contains both poles and zeros.}}
\acro{IL}{Iterated Logarithm\acroextra{, the logarithm of the logarithm}}
\acro{ILAH}{Iterated Logarithm Amplitude Histogram\acroextra{ clipping detection method}}
\acro{ILD}{Interaural Level Difference}
\acro{IMCRA}{Improved Minima Controlled Recursive Averaging\acroextra{. A technique for blindly estimating the spectrum of additive noise in a signal.}}
\acro{IMD}{Inter-Modulation Distortion}
\acro{IMSI}{International Mobile Subscriber Identity}
\acro{IMU}{Inertial Measurement Unit}
\acro{INMD}{In-service Non-intrusive Measurement Device}
\acro{INTERSPEECH}{Annual Conference of the \ac{ISCA}}
\acro{IO}{Infinitely Often}
\acro{IP}{Internet Protocol}
\acro{IPA}{International Phonetic Association}
\acro{IRS}{Inverse repeated Sequence\acroextra{. A pseudo random sequence used for impulse response measurement.}}
\acro{IRS2}[IRS]{Intermediate Reference System}
\acro{ISCA}{International Speech Communication Association}
\acro{ISDN}{Integrated Services Digital Network}
\acro{ISFT}{Inverse \acl{SFT}}
\acro{ISO}{International Organization for Standardization}
\acro{IST}{Iterative Soft Thresholding}
\acro{ISTFT}{Inverse Short Time Fourier Transform}
\acro{ITD}{Interaural Time Difference}
\acro{ITU}{International Telecommunication Union}
\acro{IUWT}{Isotropic Undecimated Wavelet (Starlet) Transform}
\acro{JADE}{Joint Approximate Diagonalization of Eigen-Matrices}
\acro{JPDA}{Joint Probabilistic Data Association}
\acro{JPEG}{Joint Photographic Experts Group}
\acro{KF}{Kalman Filter}
\acro{KL}{Karhunen-Lo{\'{e}}ve}
\acro{KLT}{Karhunen-Lo{\'{e}}ve Transform}
\acro{KST}{Kolmogorov-Smirnov Test}
\acro{LAD}{Least Absolute Deviation}
\acro{LAN}{Local Area Network}
\acro{LARS}{Least Angle Regression}
\acro{LAT}[L$_{\textrm{AT}}$]{Equivalent Continuous Sound Level\acroextra{. Also called Leq}}
\acro{LBR}{Low Bitrate Redundancy}
\acro{LC}{Local Criterion}
\acro{LCMP}{Linearly Constrained Minimum Power}
\acro{LCMV}{Linearly Constrained Minimum Variance}
\acro{LCWM}{Linear Combination of Weighted Medians}
\acro{LEM}{Loudspeaker-Enclosure-Microphone System}
\acro{Leq}[L$_{\textrm{eq}}$]{Equivalent Continuous Sound Level\acroextra{. Also called LAT}}
\acro{LF}{Liljencrants-Fant\acroextra{. The developers of a glottal waveform model}}
\acro{LHS}{Left-Hand Side}
\acro{LID}{Language Identification}
\acro{LILAH}{Least Squares Residuals Iterated Logarithm Amplitude Histogram\acroextra{ clipping detection method}}
\acro{LIME}{LInear Predictive Multi-input Equalization\acroextra{ algorithm}}
\acro{LiNoPS}{Lightweight Noise Protection System}
\acro{LLN}{Law of Large Numbers}
\acro{LLR}{Log-Likelihood Ratio}
\acro{LLS}{Logarithmic Least Squares}
\acro{LMA}{Least Mean Absolute}
\acro{LMS}{Least Mean Squares\acroextra{ adaptive filter}}
\acro{LNA}{Low Noise Amplifier}
\acro{LU}{Loudness Unit}
\acro{LUFS}{Loudness Units Full-Scale}
\acro{LOT}{Listening-Only Test}
\acro{LP}{Linear Parameter}
\acro{LP2}[LP]{Linear Predictive}
\acro{LPC}{Linear Predictive Coding\acroextra{. An autoregressive model of speech production.}}
\acro{LQO}{Listening Quality Objective}
\acro{LS}{Least Squares}
\acro{LSA}{Log Spectral Amplitude}
\acro{LSB}{Lower Side-Band}
\acro{LSB2}[LSB]{Least Significant Bit}
\acro{LSD}{Log Spectral Distortion}
\acro{LSP}{Line Spectrum Pairs}
\acro{LSR}{Late Stage Review}
\acro{LSR2}[LSR]{Least Squares Residuals\acroextra{ clipping detection method}}
\acro{LSRT}{Least Squares Residuals with Thresholding\acroextra{ clipping detection method}}
\acro{LTASS}{Long Term Average Speech Spectrum}
\acro{LTI}{Linear Time Invariant}
\acro{LTP}{Long Term Prediction}
\acro{MA}{Moving Average}
\acro{MAC}{Multiply Accumulate Operation}
\acro{MAD}{Median Absolute Deviation}
\acro{MAE}{Mean Absolute Error}
\acro{MARDY}{Multichannel Acoustic Reverberation Database at York}
\acro{MARS}{Multivariate Adaptive Regression Splines}
\acro{MCA}{Morphological Component Analysis}
\acro{MCC}{Matthew's Correlation Coefficient}
\acro{MCS}{Multidimensional Colouration Space}
\acro{MCEQ}{MultiChannel EQualisation}
\acro{MDCT}{Modified Discrete Cosine Transform}
\acro{MDL}{Minimum Description Length}
\acro{MDS}{Multidimensional Scaling}
\acro{Mel}{\acroextra{A non-uniform frequency scale corresponding to perceived frequency. It is approximately linear at low frequencies and logarithmic at high frequencies.}}
\acro{MELP}{Mixed Excitation Linear Prediction}
\acro{MFCC}{Mel-frequency Cepstral Coefficients}
\acro{MFSK}{Multi-Frequency Shift Keying}
\acro{MHT}{Multi-Hypotheses Tracking}
\acro{MI}{Mutual Information}
\acro{MIMO}{Multiple-Input-Multiple-Output}
\acro{MINT}{Multiple-input/output INverse Theorem}
\acro{MIRS}{Motorola Integrated Radio System}
\acro{MIT}{Massachusetts Institute of Technology}
\acro{MIT-LCS}{Massacchusetts Institute of Technology Laboratory for Computer Science}
\acro{ML}{Maximum Likelihood}
\acro{MLD}{Masking Level Difference}
\acro{MLS}{Maximum Length Sequence\acroextra{ of pseudo random bits.}}
\acro{MMSE}{Minimum Mean Squared Error}
\acro{MMT}{Multiscale Median Transform}
\acro{MNRU}{Modulated Noise Reference Unit}
\acro{MOM}{Mean of Maximum}
\acro{MOS}{Mean Opinion Score}
\acro{MOS-LQO}{Mean Opinion Score - Listening Quality Objective}
\acro{MP}{Matching Pursuit}
\acro{MP3}{\ac{MPEG}-2 Audio Layer III}
\acro{MPEG}{Moving Picture Experts Group}
\acro{MRF}{Markov Random Field}
\acro{MRP}{Mouth Reference Point (cf. ITU-T Rec. P.64 1999)}
\acro{MRT}{Modified Rhyme Test}
\acro{MS}{Minimum Statistics}
\acro{MSB}{Most Significant Bit}
\acro{MSC}{Mean Square Coherence}
\acro{MSN}{Multiple Subscriber Number}
\acro{MTF}{Modulation Transfer Function}
\acro{MTM}{Modified Trimmed Mean}
\acro{MUSIC}{Multiple Signal Classification}
\acro{MVDR}{Minimum Variance Distortionless Response}
\acro{NB}{Narrowband}
\acro{NCM}{Normalized Coherence Metric}
\acro{NISE}{Non-Intrusive \ac{SNR} estimation}
\acro{NISI}{Non-Intrusive Speech Intelligibility Estimation}
\acro{NISQ}{Non-Intrusive Speech Quality Estimation}
\acro{NIST}{National Institute of Standards and Technology}
\acro{NL}{Noise Level}
\acro{NLA}{Non-Linear Approximation}
\acro{NLMS}{Normalized Least Mean Squares\acroextra{ adaptive filter}}
\acro{NMCFLMS}{Normalized Multichannel Frequency Domain Least Mean Square}
\acro{NMF}{Non-negative Matrix Factorization}
\acro{NOISEX-92}{Database to Study the Effect of Additive Noise on Speech Recognition Systems}
\acro{NOIZEUS}{Noisy Speech Corpus for Evaluation of Speech Enhancement Algorithms}
\acro{NOS}{Number of Sources}
\acro{NPM}{Normalized Projection Misalignment}
\acro{NPV}{Negative Predictive Value}
\acro{NR}{Noise Reduction}
\acro{NS}{Noise Suppression}
\acro{NSRR}{Normalized Signal-to-Reverberation Ratio}
\acro{NSV}{Negative-Side Variance}
\acro{NTP}{Network Time Protocol}
\acro{OBL}{Octave Band Level}
\acro{OCA}{Open Chain Activity}
\acro{ODF}{Overdrive Factor}
\acro{OFDM}{Orthogonal Frequency Division Multiplexing}
\acro{OHC}{Outer Hair Cell}
\acro{OIM}{Objective Intelligibility Measure}
\acro{OLA}{Overlap-add}
\acro{OMP}{Orthogonal Matching Pursuit}
\acro{OSI}{Open Systems Interconnection}
\acro{OSPA}{Optimal Subpattern Assignment}
\acro{PAMS}{Perceptual Analysis Measurement System}
\acro{PARCOR}{Partial Correlation Coefficients}
\acro{PB}{Phonetically Balanced}
\acro{PBF}{Positive Boolean Function}
\acro{PCA}{Principal Components Analysis}
\acro{PCM}{Pulse-Code Modulation}
\acro{PDA}{Personal Digital Assistant}
\acro{PDE}{Partial Differential Equation}
\acro{pdf}[pdf]{Probability Density Function}
\acro{PDF}{Probability Density Function}
\acro{PE}{Parameter Estimation}
\acro{PEFAC}{Pitch Estimation Filter with Amplitude Compression}
\acro{PESQ}{Perceptual Evaluation of Speech Quality}
\acro{PF}{Psychometric Function}
\acro{pgfl}[p.g.fl.]{Probability Generating Functional}
\acro{PHAT}{Phase Transform}
\acro{PHD}{Probability Hypothesis Density}
\acro{PIV}{Pseudo-Intensity Vector}
\acro{PLC}{Packet Loss Concealment}
\acro{PLL}{Phase Locked Loop}
\acro{PM}{Phase Modulation}
\acro{PMF}{Probability Mass Function}
\acro{PMOS}{Predicted Mean Opinion Score}
\acro{POLQA}{Perceptual Objective Listening Quality Analysis}
\acro{POTS}{Plain Old Telephone Service}
\acro{PPP}{Poisson Point Process}
\acro{PPS}{Pulse-Per-Second}
\acro{PPV}{Positive Predictive Value}
\acro{PRLM}{Phoneme Recognition and Language Modelling}
\acro{PSD}{Power Spectral Density}
\acro{PSK}{Phase Shift Keying}
\acro{PSNR}{Peak Signal-to-Noise Ratio}
\acro{PSOLA}{Pitch Synchronous Overlap Add\acroextra{. A method of scaling a signal in time and pitch independently.}}
\acro{PSQM}{Perceptual Speech Quality Measurement}
\acro{PSTN}{Public Switched Telephone Network}
\acro{PWD}{Plane-Wave Decomposition}
\acro{QAM}{Quadrature Amplitude Modulation}
\acro{QMF}{Quadrature Mirror Filter}
\acro{QoE}{Quality-of-Experience}
\acro{QoS}{Quality-of-Service}
\acro{QPSK}{Quadrature Phase Shift Keying}
\acro{RASTI}{Room Acoustics Speech Transmission Index\acroextra{ (superseded by STIPA)}}
\acro{RC}{Relative Criterion}
\acro{RF}{Radio Frequency}
\acro{RFI}{Radio Frequency Interference}
\acro{RFS}{Random Finite Set}
\acro{RHS}{Right-Hand Side}
\acro{RIP}{Restricted Isometry Property}
\acro{RIR}{Room Impulse Response}
\acro{RLS}{Recursive Least Squares\acroextra{ adaptive filter}}
\acro{RLSD}{Relative Log Spectral Distortion}
\acro{RMCLS}{Relaxed Multichannel Least Squares}
\acro{RMCLS-CIT}{Relaxed MultiChannel Least-Squares with Constrained Initial Taps}
\acro{RMS}{Root Mean Square}
\acro{RMSE}{Root Mean Square Error}
\acro{ROC}{Receiver Operating Characteristic}
\acro{ROHC}{Robust Header Compression}
\acro{RPE}{Regular Pulse Excitation}
\acro{RS}{Reverberation Suppression}
\acro{RT}{Reverberation Time}
\acro{RTAN}{Robustness to Additional Noise}
\acro{RTF}{Real-Time Factor}
\acro{RTF2}[RTF]{Room Transfer Function}
\acro{RV}{Random Variable}
\acro{RVP}{Recursive Vector Projection}
\acro{SAP}{Speech And Audio Processing}
\acro{SAR}{Speech-to-Artifact Ratio}
\acro{SAR2}[SAR]{Speaker Alternation Rate}
\acro{SCAF}{Single Channel Adaptive Filter}
\acro{SCB}{Stochastic Codebook}
\acro{SCOT}{Smoothed Coherence Transform}
\acro{SC-PHD}{Single Cluster \ac{PHD}}
\acro{SCR}{Signal-to-Competition Ratio}
\acro{SCRIBE}{Spoken Corpus of British English}
\acro{SCT}{Speech Corruption Toolkit}
\acro{SCT2}{Short Conversation Test}
\acro{SD}{Semantic Differential}
\acro{SDB}{Superdirective Beamformer}
\acro{SDD}{Spectral Decay Distributions}
\acro{SDDMSB}{\ac{SDD} with Mel-spaced frequency bands}
\acro{SDDSA}{\ac{SDD} with Mel-spaced frequency bands and selective averaging}
\acro{SDR}{Software Defined Radio}
\acro{SDR2}{Speech}
\acro{SDRAM}{Synchronous Dynamic Random Access Memory}
\acro{SDT}{Speech Description Taxonomy}
\acro{SEMG}{Surface Electromyography}
\acro{SFDR}{Spurious Free Dynamic Range}
\acro{SFT}{Spherical Fourier Transform}
\acro{SH}{Spherical Harmonic}
\acro{SHD}{Spectral Harmonic Decomposition}
\acro{SHD2}[SHD]{Spherical Harmonic Domain}
\acro{SIE}{System Identification Error}
\acro{SII}{Speech Intelligibility Index}
\acro{SIImod}{Speech Intelligibility Index in the modulation domain}
\acro{SIM}{Subscriber Identity Module}
\acro{SIMO}{Single-Input-Multiple-Output}
\acro{SINAD}{Signal-to-Noise and Distortion Ratio}
\acro{SIP}{Session Initiation Protocol}
\acro{SIR}{Signal-to-Interference Ratio}
\acro{SIR2}[SIR]{Sequential Importance Resampling}
\acro{SIREAC}{Simulation of REal Acoustics\acroextra{ Software Tool}}
\acro{SIS}{Sequential Importance Sampling}
\acro{SL}{Speech Level}
\acro{SLAM}{Simultaneous Localisation and Mapping}
\acro{SLLN}{Strong Law of Large Numbers}
\acro{SLM}{Sound Level Meter}
\acro{SMA}{Spherical Microphone Array}
\acro{SMERSH}{Spatiotemporal Averaging Method for Enhancement of Reverberant Speech}
\acro{SMIR}{Spherical Microphone array Impulse Response}
\acro{SMPTE}{Society of Motion Picture and Television Engineers}
\acro{SMS}{Short Message Service}
\acro{SNR}{Signal-to-Noise Ratio}
\acro{SNR2}[SNR]{Speech-to-Noise Ratio}
\acro{SNT}{Subspace Noise Tracking\acroextra{ algorithm}}
\acro{SOBM}{STOI-optimal Binary Mask}
\acro{SOLA}{Synchronous Overlap Add\acroextra{. A method of scaling a signal in time and pitch independently.}}
\acro{SPC}{Specificity}
\acro{SPEECON}{Speech Databases for Consumer Devices}
\acro{SPHERE}{NIST SPeech Header REsources\acroextra{ software with embedded Shorten Compression}}
\acro{SPIN}{Speech Perception In Noise}
\acro{SPL}{Sound Pressure Level}
\acro{SPP}{Speech Presence Probability}
\acro{SPQA}{Speech Quality Assurance Package}
\acro{SQNR}{Signal-to-Quantization Noise Ratio}
\acro{SR}{Sparse Representation}
\acro{SRA}{Statistical Room Acoustics}
\acro{SRI}{SRI International\acroextra{. Formerly Standford Research Institute}}
\acro{SRMR}{Speech-to-Reverberation Modulation Energy Ratio}
\acro{SRP}{Steered Response Power}
\acro{SRR}{Signal-to-Reverberation Ratio}
\acro{SRT}{Speech Reception Threshold\acroextra{ also known as Speech Recognition Threshold}}
\acro{SS}{Spectral Subtraction}
\acro{SSB}{Single Side-Band}
\acro{SSI}{Synthetic Sentence Identification}
\acro{SSL}{Sound Source Localization}
\acro{SSN}{Simultaneous Switching Noise}
\acro{SSOBM}{Stochastic STOI-optimal Binary Mask}
\acro{SSW}{Staggered Spondaic Word}
\acro{STFT}{Short Time Fourier Transform}
\acro{STI}{Speech Transmission Index}
\acro{STIPA}{Speech Transmission Index for Public Address Systems}
\acro{STITEL}{Speech Transmission Index for Telecommunication Systems}
\acro{STMI}{Spectro-Temporal Modulation Index}
\acro{STNR}{\ac{NIST}'s Speech-to-Noise Ratio\acroextra{ Estimation Algorithm}}
\acro{STOI}{Short-Time Objective Intelligibility Measure}
\acro{STQ}{Speech Processing, Transmission and Quality Aspects}
\acro{STSA}{Short Time Spectral Analysis}
\acro{STSA1}[STSA]{Short Time Spectral Amplitude}
\acro{SUS}{Semantically Unpredictable Sentences}
\acro{SVD}{Singular Value Decomposition}
\acro{SVM}{Support Vector Machine}
\acro{T20}[$T_\textrm{20}$]{Reverberation Time\acroextra{ to decay by $20$ dB}}
\acro{T30}[$T_\textrm{30}$]{Reverberation Time\acroextra{ to decay by $30$ dB}}
\acro{T60}[$T_\textrm{60}$]{Reverberation Time\acroextra{ to decay by $60$ dB}}
\acro{TBM}{Target Binary Mask}
\acro{TDHS}{Time Domain Harmonic Scaling\acroextra{. A method of scaling a signal in time and pitch independently.}}
\acro{TDOA}{Time-Difference-of-Arrival}
\acro{TDT}{Tone Decay Test}
\acro{TF}{Time-Frequency}
\acro{TFGM}{Time-Frequency Gain Modification\acroextra{. An approach to signal enhancement in which a signal is multiplied by a gain function in the time-frequency domain.}}
\acro{THD}{Total Harmonic Distortion}
\acro{TI}{Texas Instruments, Inc.}
\acro{TIMIT}{\ac{TI}-\ac{MIT} speech corpus}
\acro{TIPHON}{Telecommunication and Internet Protocol Harmonization Over Networks}
\acro{TLS}{Total Least-Squares}
\acro{TN}{True Negative}
\acro{TNR}{True Negative Rate}
\acro{TOA}{Time-of-Arrival}
\acrodefplural{TOA}[TOAs]{Times-of-Arrival}
\acro{TOSQA}{Telekom Objective Speech Quality Assessmentt}
\acro{TP}{True Positive}
\acro{TP2}[TP]{Trivial Pursuit}
\acro{TPCC}{Trivial Pursuit with Clipping Constraints}
\acro{TPR}{True Positive Rate}
\acro{TSE}{Taylor Series Expansion}
\acro{TVAR}{Time-varying Autoregression}
\acro{UDP}{User Datagram Protocol}
\acro{UHF}{Ultra High Frequency}
\acro{UKF}{Unscented Kalman Filter}
\acro{ULA}{Uniform Linear Array}
\acro{ULF}{Ultra Low Frequency}
\acro{UMTS}{Universal Mobile Telecommunications Service}
\acro{US}{United States}
\acro{UTBM}{Universal Target Binary Mask}
\acro{VAD}{Voice Activity Detector}
\acro{VBR}{Variable Bit-Rate}
\acro{VCV}{Vowel-Consonant-Vowel}
\acro{VGC}{Voice Grade Channel}
\acro{VoIP}{Voice Over Internet Protocol}
\acro{VRT}{Vlaamse Radio- en Televisieomroeporganisatie\acroextra{. (Flemish Radio and Television Broadcasting Organization)}}
\acro{VSELP}{Vector Sum-excited Linear Prediction}
\acro{VST}{Virtual Studio Technology\acroextra{. An interface standard developed by Steinberg for adding plugins to an audio editor.}}
\acro{WADA}{Waveform Amplitude Distribution Analysis}
\acro{WASPAA}{IEEE Workshop on Applications of Signal Processing to Audio and Acoustics}
\acro{WAV}{Waveform Audio File Format}
\acro{WAVE}{Waveform Audio File Format}
\acro{WB}{Wideband}
\acro{WER}{Word Error Rate}
\acro{WGN}{White Gaussian Noise}
\acro{WLAN}{Wireless \ac{LAN}}
\acro{WLLN}{Weak Law of Large Numbers}
\acro{WMA}{Windows Media Audio}
\acro{WNG}{White Noise Gain}
\acro{ZOS}{Zero-Order Statistics}

\acro{CC}{Cepstral Coefficients}
\acro{LDA}{Linear Discriminant Analysis}
\acro{LFCC}{Linear-frequency Cepstral Coefficients}
\acro{MRI}{Magnetic Resonance Imaging}
\acro{MVN}{multivariate normal}
\acro{OA}{Osteoarthritis}
\acro{OCA}{Open Chain Activity}
\acro{PAG}{Phonoarthrography}
\acro{QDA}{Quadratic Discriminant Analysis}
\acro{VAG}{Vibroarthrography}
\acro{kNN}{k-Nearest Neighbors}
\end{acronym}

\maketitle

%
%
\begin{abstract}
\textit{ Objective:} In this work the potential of non-invasive detection of knee osteoarthritis is investigated using the sounds generated by the knee joint during walking. \textit{Methods:} The information contained in the time-frequency domain of these signals and its compressed representations is exploited and their discriminant properties are studied. Their efficacy for the task of normal vs abnormal signal classification is evaluated using a comprehensive experimental framework. Based on this, the impact of the feature extraction parameters on the classification performance is investigated using \acl{CART}s, \acl{LDA} and \acl{SVM} classifiers. \textit{Results:} It is shown that classification is successful with an area under the \acl{ROC} curve of 0.92. \textit{Conclusion:} The analysis indicates improvements in classification performance when using non-uniform frequency scaling and identifies specific frequency bands that contain discriminative features. \textit{Significance:} Contrary to other studies that focus on sit-to-stand movements and knee flexion/extension, this study used knee sounds obtained during walking. The analysis of such signals leads to non-invasive detection of knee osteoarthritis with high accuracy and could potentially extend the range of available tools for the assessment of the disease as a more practical and cost effective method without requiring clinical setups.
%
\end{abstract}

\begin{IEEEkeywords}
knee joint sounds, walking, osteoarthritis, time-frequency analysis, pattern classification
\end{IEEEkeywords}

%
\IEEEpeerreviewmaketitle

%
%
\section{Introduction}
\label{intro}
%
%
%
%
%
\IEEEPARstart{O}{steoarthritis} is the most common disabling and financially burdensome of all musculoskeletal diseases, and prevalence is rising. It occurs most frequently in the knee, affecting 1 in 5 adults over the age of 45~\cite{Arthritis2013}. It leads to pain, stiffness and swelling of the joint, greatly degrading the quality of life. Risk of \ac{OA} is associated with increased mechanical wear, such as through older age and high body weight~\cite{Lane2002}. Currently, there is no cure and treatments aim to manage symptoms through lifestyle modification, physio- and pharmacological therapy~\cite{Arthritis2013}. In severe cases, total knee replacement is required.

%
Clinical detection of knee \ac{OA} relies on a combination of patient reported symptoms and medical imaging of cartilage and subchondral bone degradation. Current imaging methods such as X-ray, \ac{MRI} and ultrasound have poor sensitivity in early disease and as a result, at the time of diagnosis, \ac{OA} is already at a progressed stage, and understanding of its cause and development is still limited. Additionally, current imaging techniques provide images of the static anatomical structure of knee joints at a particular posture and are therefore limited in assessing the dynamic integrity of the knee during a dynamic \ac{OCA}, for example when the foot leaves and makes contact again with the ground as happens during walking. This is important since \ac{OA} patients experience pain and discomfort when their knee is functional. Although dynamic \ac{MRI} produces good measurements in the assessment of knee function, it is normally not practical in terms of cost and accessibility~\cite{Gold2003}. 
Hence, there exists the need for a quick, non-invasive, portable and cheaper technique that would ideally be accessible in a non-clinical environment and could be used as a screening tool for the mild disease cases.

Joints generate sounds during movement. When the knee is active, the joint between the tibia and the femur bones moves. The regions and perhaps the quality of joint surfaces coming into contact are different at each angular position, generating therefore a number of different sounds during movement. 
In 
healthy 
knee joints, the bones have smooth surfaces due to a thin layer of cartilage and are separated by a protective space filled with synovial fluid to reduce friction~\cite{Welsh1980}. They are able to move freely and the level of sound emitted is low. 
In \ac{OA} knees this structure is degraded and the protective space and associated lubrication reduce, resulting in increased friction which accelerates the wear of cartilage~\cite{Lane2002}. This increased friction makes the knee more noisy during motion.

The potential for using knee joint sounds for diagnostic purposes has been known for many years. Blodgett, in 1902, reported on auscultation of the knee, with attention to sounds of normal joints and their change with repetitive motion, where a relation between an increase of sound activity and age was noted~\cite{Blodgett1902}. In 1913, Bircher reported that different types of meniscal injury generate distinctive sound signals~\cite{Bircher1913,Tavathia1992}. Steindler in 1937, used a system consisting of a cardiophone, an oscilloscope and a recorder to study 397 knees~\cite{Steindler1937}. He found a relation between pathologies and the pitch, amplitude and the sequence of sounds and was able to classify the joints based on these features. It was observed, however, that it was difficult to separate other body sounds such as muscle activity from articular cartilage sounds. In~\cite{Fischer1960} the authors claim that sounds could be detected in rheumatoid arthritis before any changes were observable in an X-Ray image but no further work was conducted to confirm this claim. 

Auscultation based \ac{PAG} utilises microphones in the audible frequency range to record sounds generated during movement. Important work on \ac{PAG} by Chu et al. reported that the spectral activity of pathological knees (recorded during active motion) spanned the entire audible frequency range and the signals' acoustic power increased with severity of cartilage damage~\cite{Chu1976,Chu1976a,Chu1977,Chu1978}. Significant work was directed to the development of \ac{VAG} as an alternative to \ac{PAG} which relies on accelerometer sensors, operating at frequencies below 1~kHz, to pick up mechanical vibrations. Algorithms proposed for classifying the knee \ac{VAG} signals according to pathological conditions, range from linear prediction modelling~\cite{Rangayyan1997,Krishnan1997} to time-frequency analysis~\cite{Krishnan2000,Kim2008b,Kim2009a} and wavelet matching pursuit decomposition \cite{Wu2009,Cai2013}. Several features have been used for classification, including spectrogram features, waveform variability parameters, statistical features~\cite{Rangayyan2008}, fundamental frequency, mean amplitude of pitch and their jitter and shimmer~\cite{Kim2006b,Kim2012}. Classifiers used in the literature range from early neural network architectures~\cite{Kim2009a,Rangayyan2009} to maximal posterior probability decision criterion~\cite{Wu2013b}, bagging ensemble and multiple classifier system based on adaptive weighted fusion~\cite{Wu2009}. A thorough description of \ac{VAG} analysis can be found in~\cite{Wu2015}.

The use of \ac{AE1} at ultrasonic frequencies was explored as a potential biomarker for assessing the knee joint condition. In~\cite{Mascaro2009} piezoelectric contact sensors were used to capture ultrasonic \ac{AE1} signals (50~kHz to 200~kHz) emitted during sit-to-stand movements and was demonstrated, using Principal Component Analysis, that healthy and \ac{OA} knees are separable in the feature space. It was further concluded that \ac{OA} knees produce substantially more \ac{AE1} events with higher peak magnitude and average signal level~\cite{Shark2010,Shark2011,Chen2011}. 

\ac{AE1} analysis during knee flexion-extension was also explored in the context of knee injury rehabilitation~\cite{Toreyin2016a,Hersek2018,Inan2018}. 
%
In~\cite{Hersek2018}, a 64-dimensional feature representation of 200~ms frames of the knee sound signal was used, from which a \ac{kNN} graph was constructed. 
A graph based metric was then proposed to quantify the homogeneity of the feature matrix without modelling the underlying distribution. Based on this metric, it was concluded that injured knee joints produce more heterogeneous features than healthy knee joints~\cite{Hersek2018}. Although this approach alleviates the need for prior algorithm training, it is only accurate when sound data from both knees of an injured subject is available since the study focused on the intra-subject knee sound differences. Inter-subject sound differences were not considered. This is one of the most challenging aspects of knee joint sound analysis as there is strong variability in the knee sounds amongst individuals which is likely due to their joints' structural differences~\cite{Inan2018}. 

%
%
The movement protocols most often reported in the literature, for \ac{OA} and other arthritis related studies, are knee flexion-extension and sit-to-stand movements~\cite{Wu2015,Mascaro2009,Krishnan2001}. 
In the work presented here, acoustic signals are captured by a contact microphone attached to the patella, while patients are walking on a specialised treadmill. Various feature-based descriptions for these signals are investigated. 
In particular, discriminative features are sought that are relevant to the analysis and classification of normal (clinically healthy) and abnormal (clinically \ac{OA}) knee joints. 
%
A preliminary version of this work was presented in~\cite{Yiallourides2018}. Here this work is extended to consider time-frequency representations of the knee acoustic signals as features and examine their discriminatory power. In addition, a study is presented on the effect on the classification performance of the choice of parameters in the feature extraction step.
%

The main aim of this work 
is 
to answer four questions: (a)~Does the classification performance improve when the \ac{DFT} spectrum is compressed using triangular filter-banks? (b)~Does the classification performance improve when the natural logarithm and \ac{DCT} are used instead of only the \ac{DFT}? (c)~Is the classification performance better when using uniform or non-uniform frequency spacing in the analysis? 
(d)~Which frequency ranges of \ac{AE1} signals contain more discriminative information and hence are important for \ac{OA} classification? In answering these questions, an insight will be obtained into which features best characterise normal and \ac{OA} knees. Several classifiers are used but their optimisation is beyond the scope of this paper.

The remainder of the paper is structured as follows. Section \ref{sec:featExtSection} describes the features considered in this work and introduces the notation that is used throughout the paper. This is followed by feature analysis and selection in Section \ref{sec:analAndSelectSection}. Experiments along with results and discussion are presented in Section \ref{sec:ChapSegmClasifExp} 
where detailed information about the data acquisition and assessment protocol can be found in~\ref{dataAcq}. Finally, Section~\ref{conclusions} concludes the paper with a summary of the proposed work.

%
%
\section{Feature extraction}
\label{sec:featExtSection}
Acoustic signals are recorded over the patella using a contact microphone, as will be described in Section~\ref{dataAcq}. Let $\gls*{iKneeSig}$ denote the signal at discrete time index $\gls*{DTindex}$ captured by the patella microphone for the $i^{\text{th}}$ knee in the data-set, where $i=1,2,\ldots,I$ for $I$ knees. Prior to extracting features, all recorded signals are normalised to have equal \ac{RMS} level. 
We hypothesize that the acoustic artifacts caused by walking on the treadmill are uncorrelated with the features used for the analysis. Furthermore, it is assumed that sounds related to abnormalities appear within time periods of $\gls*{assumpWindow}$~seconds. Accordingly, $\gls*{iKneeSig}$ is divided into non-overlapping segments of length $\gls*{assumpWindow}$, denoted as $\gls*{iKneeSigjSegm}$ for $j=1,2,\ldots,J_i$ segments. Each segment is then labelled for classification according to the condition of the knee from which it was obtained.


A signal segment $\gls*{iKneeSigjSegm}$ is further divided into frames of length $\gls*{genFrameSizems}$~ms with 50\% overlap. This creates an $\gls*{numFrames}\times \gls*{genFrameSizeSamps}$ matrix $\gls*{enframedSigMat}$ where $\gls*{numFrames}$ is the number of frames and $\gls*{genFrameSizeSamps}$ is the frame length in samples. Considering a hanning window of length $\gls*{genFrameSizeSamps}$ transformed into the diagonal matrix $\gls*{hanningDiagMat}$, the \ac{DFT} of $\gls*{enframedSigMat}$ can be computed as 
\begin{equation}
\boldsymbol {\Psi_f}= (\gls*{enframedSigMat}\gls*{hanningDiagMat})\gls*{dftMatCoefs}
\end{equation}
where

\begin{equation*} 
\gls*{dftMatCoefs} = 
\begin{bmatrix}
1		 & 1 	& 1	    & \dots 	& 1 \\
1	     & e^{ -\frac{2\gls*{pi}\gls*{imagI}}{\gls*{genFrameSizeSamps}} } &e^{ -\frac{4\gls*{pi}\gls*{imagI}}{\gls*{genFrameSizeSamps}} }  &\dots 	&e^{ -\frac{2\gls*{pi}\gls*{imagI}(\gls*{genFrameSizeSamps}-1)}{\gls*{genFrameSizeSamps}} } \\
\vdots	 & \vdots  	 & \vdots   	& \ddots 	& \vdots \\
1	     & e^{ -\frac{2\gls*{pi}\gls*{imagI}(\gls*{genFrameSizeSamps}-1)}{\gls*{genFrameSizeSamps}} } &e^{ -\frac{4\gls*{pi}\gls*{imagI}(\gls*{genFrameSizeSamps}-1)}{\gls*{genFrameSizeSamps}} }  &\dots 	&e^{ -\frac{2\gls*{pi}\gls*{imagI}(\gls*{genFrameSizeSamps}-1)(\gls*{genFrameSizeSamps}-1)}{\gls*{genFrameSizeSamps}} } \\
\end{bmatrix}
\end{equation*}
is the Vandermonde matrix for the roots of unity, otherwise known as the DFT matrix in this context. Each element of $\gls*{dftMatCoefs}$ is given by $ e^{ -\frac{2\gls*{pi}\gls*{imagI}\gls*{DTindex}\gls*{freqindex}}{\gls*{genFrameSizeSamps}} }$ where for each row ${\gls*{DTindex}=0,1,\ldots,\gls*{genFrameSizeSamps}-1}$ and for each column ${\gls*{freqindex}=0,1,\ldots,\gls*{genFrameSizeSamps}-1}$ where $\gls*{freqindex}$ is the frequency index. By taking the magnitude of each element in $\boldsymbol {\Psi_f}$ and retaining only the first  ${K = \text{floor}(1+\gls*{genFrameSizeSamps}/2)}$ columns, the matrix $\gls*{enframedSigMatDft}$ is constructed.

A filter-bank with $\gls*{numFilters}$ triangular band-pass filters 
linearly spaced in frequency is used to construct the matrix
\begin{equation*}
\gls*{linFiltbankMat} = 
\begin{bmatrix}
U_1(0)				& U_2(0) 		    & \dots 	& U_{\gls*{numFilters}}(0) \\
U_1(\frac{2\pi}{K})	& U_2(\frac{2\pi}{K})& \dots 	& U_{\gls*{numFilters}}(\frac{2\pi}{K}) \\
\vdots	    		& \vdots  	    	& \ddots 	& \vdots \\
U_1(\frac{2\pi(K-1)}{K})	& U_2(\frac{2\pi(K-1)}{K})& \dots 	& U_{\gls*{numFilters}}(\frac{2\pi(K-1)}{K}) \\
\end{bmatrix}
\end{equation*}
where each element is the magnitude of the bandwidth of a single filter at a single frequency bin $k=0,1,\ldots,K-1$. A matrix $\gls*{melFiltbankMat}$ is similarly constructed from triangular filters that are equally spaced along the mel-frequency axis which is defined as in~\cite{Makhoul1976}. A compact spectrum representation can then be obtained as
\begin{align}
\gls*{LinDftMat} &= \gls*{enframedSigMatDft}\gls*{linFiltbankMat}\\
\gls*{MelDftMat} &= \gls*{enframedSigMatDft}\gls*{melFiltbankMat}~.
\end{align}
The resultant matrices are $\gls*{numFrames}\times \gls*{numFilters}$ and in this way dimensionality reduction is achieved.

The columns of $\gls*{LinDftMat}$, $\gls*{MelDftMat}$ and $\gls*{enframedSigMatDft}$ can be considered as distributions of the spectrum magnitude values for particular frequency bands for the former two and bins for the latter. 
From each such distribution a feature vector $\boldsymbol{f}$ is extracted using 11 statistical parameters that capture certain signal attributes that aim to highlight differences between \ac{OA} and healthy signals. These parameters are chosen in this work to be the mean, kurtosis, variance, skewness, max, min and the $10^\text{th}$, $25^\text{th}$, $50^\text{th}$, $75^\text{th}$, $90^\text{th}$ percentiles. The vector $\boldsymbol{f}$ can be obtained by defining an operator $\gls*{statExtractOper}(\cdot)$ which acts on a matrix and returns the 11 statistical parameters of each column. Therefore,
\begin{align*}
\gls*{LinFreqSTFTfeatVec} &= \gls*{statExtractOper}(\gls*{LinDftMat}) = [\boldsymbol{f}_1^{E},\boldsymbol{f}_2^{E},\ldots,\boldsymbol{f}_{\gls*{numFilters}}^{E}]^T \\ \numberthis
\gls*{MelFreqSTFTfeatVec} &= \gls*{statExtractOper}(\gls*{MelDftMat}) = [\boldsymbol{f}_1^{D},\boldsymbol{f}_2^{D},\ldots,\boldsymbol{f}_{\gls*{numFilters}}^{D}]^T \\ 
\gls*{STFTfeatVec} &= \gls*{statExtractOper}(\gls*{enframedSigMatDft}) = [\boldsymbol{f}_1^{F},\boldsymbol{f}_2^{F},\ldots,\boldsymbol{f}_{K}^{F}]^T ~.
\end{align*} 
\begin{figure*}[t]
	\centering
	\includegraphics{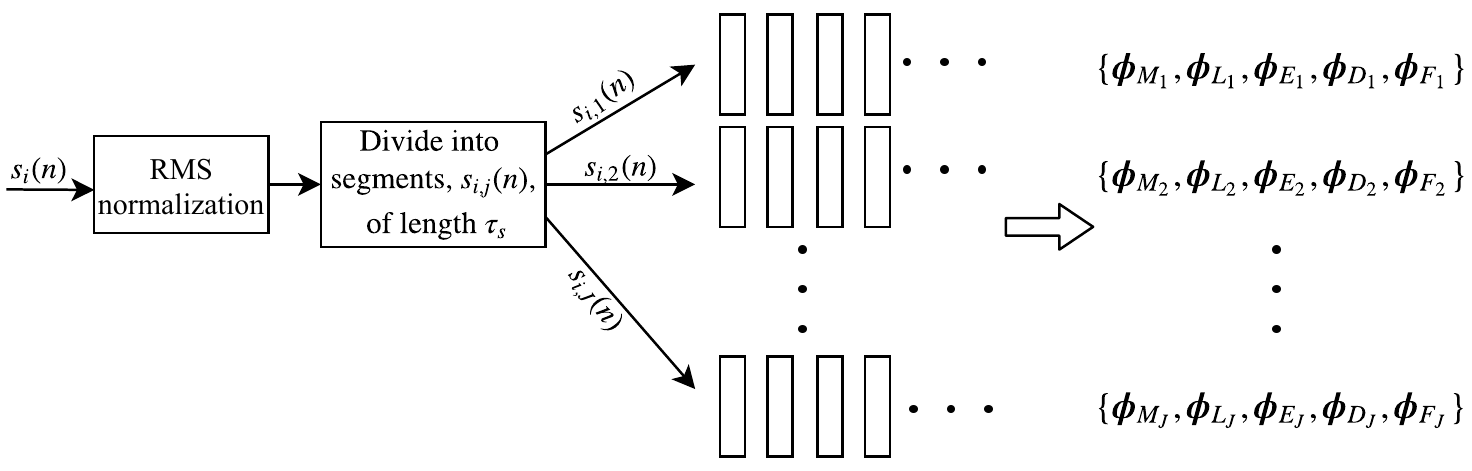}
	\centering
	\includegraphics{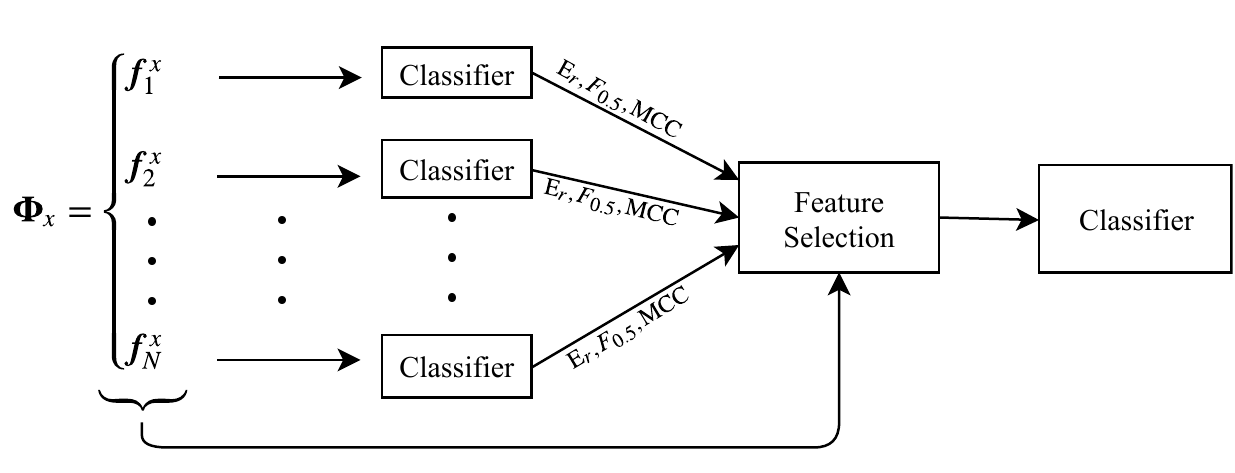}
	\caption{\textbf{Top:} Feature extraction process for a single $s_i(n)$ that is divided into $J_i$ segments. \textbf{Bottom:} Feature analysis and subset selection process with subsequent classification presented for feature set $\boldsymbol{\Phi}_x$ where $x$ denotes any of the symbols $\{D,E,F,L,M\}$ and $N=K$ for ${x:=F}$ or $N=N_B$ otherwise.\label{fig:journal_methodology}}
\end{figure*}
Extraction procedures inspired by human auditory perception are widely used in many applications. \ac{MFCC} are a common choice of features that are successfully used in speech recognition and music genre classification. 
%
\ac{MFCC} have been previously used for the analysis of \ac{VAG} signals,~\cite{Seng2013a,Seng2013b}, but have not so far been used for \ac{OA} detection from the analysis of acoustic signals emitted from the knee and sensed at the patella. 
Their extraction process involves mapping the power of the \ac{STFT} spectrum using triangular overlapping windows onto the mel scale which is designed to approximate the human auditory system's response. The aim is to exploit the property of the mel scale and apply it to the knee signals. In particular, for the sounds heard as pops, clicks, grindings etc. during knee motion. The use of the mel scale for knee signals is motivated by the fact that these sounds can be distinguished in a recording by even the untrained ear with minimal effort and that they are sounds that are likely generated by the friction between the tibia and the femur bones which in turn is caused by the effects of \ac{OA} in the knee.

\ac{MFCC} are computed according to \cite{Quatieri2002} and in the same way, but replacing the mel-frequency filter-bank with linearly spaced filters, a set of \ac{LFCC} is also computed. Hence
\begin{align*}
\label{equ:MLfccMats}
\gls*{mfccPsiD} &= \gls*{DCToper}( \log (\gls*{MelDftMat}) ) \\ \numberthis
\gls*{LfccPsiE} &= \gls*{DCToper}( \log (\gls*{LinDftMat}) )
\end{align*}
denote the two sets of \ac{CC}, where $\gls*{DCToper}(\cdot)$ is the \ac{DCT} operator. Applying the same thinking as with the \ac{STFT} based representations, the $\gls*{statExtractOper}(\cdot)$ operator is employed to obtain the statistical representation of the matrices in (\ref{equ:MLfccMats}) as
\begin{align*}
\gls*{mfccfeatVec} &= \gls*{statExtractOper}(\gls*{mfccPsiD}) = [\boldsymbol{f}_1^{M},\boldsymbol{f}_2^{M},\ldots,\boldsymbol{f}_{\gls*{numFilters}}^{M}]^T \\ \numberthis
\gls*{LfccfeatVec} &= \gls*{statExtractOper}(\gls*{LfccPsiE}) = [\boldsymbol{f}_1^{L},\boldsymbol{f}_2^{L},\ldots,\boldsymbol{f}_{\gls*{numFilters}}^{L}]^T ~.
\end{align*}

Given the fixed frame segmentation process employed using short time-frames, it is likely that a knee sound related to \ac{OA} might extend to more than one frame. By taking the time derivatives of the coefficients, the information present in the evolution of these sounds across a multiple of frames can be extracted. Therefore, for each element in $\gls*{MelDftMat}$, $\gls*{LinDftMat}$, $\gls*{mfccPsiD}$ and $\gls*{LfccPsiE}$, defined as a static coefficient and denoted as $a_t$, the 
first derivatives in time are computed using
\begin{equation}
\label{equ:derivative}
d_t=\frac{\sum\limits_{u=1}^{U}u(a_{t+u}-a_{t-u})}{2\sum\limits_{u=1}^{U}u^2}
\end{equation}
where $U=4$ and $d_t$ is termed a delta coefficient from frame $t$ computed using the static coefficient of that frame. The second derivatives, termed as delta-delta coefficients, are also computed using (\ref{equ:derivative}) but replacing $a_t$ with $d_t$ and setting $U=1$ which makes it a simple difference equation. The choice of $U=1$ and $U=4$ is adopted from speech recognition.

All the values of the first derivative obtained from a single feature in either $\gls*{mfccPsiD}, \gls*{LfccPsiE}, \gls*{LinDftMat}$ and $\gls*{MelDftMat}$, (i.e. cepstral coefficient or frequency band), can be considered to form a distribution from which the 11-dimensional vector $\boldsymbol{f}$ is extracted. The same is performed for all the features in these 4 matrices for their first and second derivatives and the new vectors are appended to the appropriate $\boldsymbol{\phi}$ feature matrix. Performing the above process for all signal segments produces 5 feature sets that will be used for the classification experiments. These are
\begin{align*}
\label{featspaces}
\gls*{LinFreqSTFTfeatSet} &= [\boldsymbol{\phi_{E}}_1, \boldsymbol{\phi_{E}}_2,\ldots,\boldsymbol{\phi_{E}}_c] \\
\gls*{MelFreqSTFTfeatSet} &= [\boldsymbol{\phi_{D}}_1, \boldsymbol{\phi_{D}}_2,\ldots,\boldsymbol{\phi_{D}}_c] \\ \numberthis
\gls*{STFTfeatSet} &= [\boldsymbol{\phi_{F}}_1, \boldsymbol{\phi_{F}}_2,\ldots,\boldsymbol{\phi_{F}}_c] \\ 
\gls*{LfccfeatSet} &= [\boldsymbol{\phi_{L}}_1, \boldsymbol{\phi_{L}}_2,\ldots,\boldsymbol{\phi_{L}}_c] \\
\gls*{mfccfeatSet} &= [\boldsymbol{\phi_{M}}_1, \boldsymbol{\phi_{M}}_2,\ldots,\boldsymbol{\phi_{M}}_c] 
\end{align*}
for $c=\sum_{i=1}^{I}J_i$ total segments.
In the following Section, the effectiveness of the signal segment parameterisation using the 11 statistics is examined and their discriminant power for the task of normal vs abnormal segment classification is studied.

%
%
\section{Feature Analysis and Selection}
\label{sec:analAndSelectSection}
Let $x$ refer to any of the symbols $\{D,E,F,L,M\}$. The aim is to evaluate the discriminant power of each $\boldsymbol{f}_{i}^{x}~{\forall~i=1,\ldots,N}$ in (\ref{featspaces}) independently, where $N=K$ for ${x:=F}$ or ${N=\gls*{numFilters}}$ otherwise. The classifier employed for this purpose is \ac{SVM} because it is efficient for small training data and avoids making any assumptions on the underlying data distribution~\cite{Cristianini2000}. This makes it a suitable choice since the distribution is unknown. The linear kernel \ac{SVM} is used, denoted as $\text{SVM}_{l}$, as it is the simplest form of kernel and is less prone to overfitting than other more complex kernels~\cite{Cristianini2000}.

The knee database, obtained as described in Section~\ref{dataAcq} and used in the experiments to follow, is comprised of 19 normal knees and 21 abnormal from which the signal segments $\gls*{iKneeSigjSegm}$ are obtained using a window size of $\gls*{assumpWindow}$ seconds. A cross-validation procedure is employed using 5 groups, randomly constructed from the database, with a normal to abnormal knees ratio of 3:5, 3:5, 3:5, 5:3, 5:3 for each group which are then made up with the segments of their constituting knees. In this way the problem of having segments of a knee signal in more than one group is avoided. Some variability in the group sizes exists given that some knee recordings in the database are longer than others and therefore have more segments. Four groups are used for training the \ac{SVM} model which is then tested on the group left out. This is repeated until all 5 groups are evaluated. Prior to this, the training data (4 groups) is scaled by subtracting the mean and normalising by the variance. The same scale values are then applied to the test data (the group left out). The above procedure is executed 100 times in order to reduce the variance of the estimator and the results are averaged at the end.

The performance of each feature is assessed based on several metrics. Relying only on the error rate $(\gls*{errRate})$ is often not sufficient to draw safe conclusions since potential classification errors other than the number of misclassified observations are not captured by the error rate. Hence, the $\gls*{f05score}$ measure (a variation of $\gls*{f1score}$) and \ac{MCC} are also used. Both are computed from the confusion matrix~\cite{Powers2011}. From a clinical perspective, the false prediction of abnormal segments as normal is worse than the contrary. $\gls*{f05score}$ emphasises this error type more than $\gls*{f1score}$ and is thus preferred. \ac{MCC} is a balanced measure ranging from -1 (prediction totally different from observation), to 1 (perfect prediction), with 0 stating no better than random prediction~\cite{Baldi2000}.

%
Following the pre-processing with feature extraction and analysis steps, 
the selection of feature subsets for subsequent classification and analysis is performed. The selection method used is a hybrid of a filter and a wrapper approach. First, the features are ranked in each of the 3 metric categories (the filter step) and with the application of thresholds, $\big[\gls*{errRateThres},\gls*{f05scoreThres},\gls*{mccThres}\big]$ for $\big[\gls*{errRate},\gls*{f05score},\text{MCC}\big]$ respectively, the best \textit{N} features are selected, where \textit{N} is dictated by those that satisfy all thresholds. 
Secondly, by allowing $\gls*{errRateThres}$, $\gls*{f05scoreThres}$ and $\gls*{mccThres}$ to vary in the range $\big[0,1\big]$ with discrete steps of size $w$, the entire feature space is searched and all possible subsets are constructed subject to these constraints (the wrapper step). The three metrics are bounded in a continuous range and therefore defining a discrete set of constraints is necessary in order to make the search space tractable as it is not practical to test all possible combinations of features as an exhaustive search. 
%
%
This feature selection method forms nested subset of features,
\begingroup\makeatletter\def\f@size{8.5}\check@mathfonts
\def\maketag@@@#1{\hbox{\m@th\large\normalfont#1}}%
{\begin{IEEEeqnarray*}{rCl}
S_{1}^{x}&=\{\boldsymbol{f}_q^{x},...,\boldsymbol{f}_{q+N_1}^{x}\} \textbf{ s.t. } \theta_{er}^{1}\geq\{J(\boldsymbol{f}_q^{x}),...,J(\boldsymbol{f}_{q+N_1}^{x})\} \geq\{\theta_{0.5}^{1},\theta_{mcc}^{1}\}\\
S_{2}^{x}&=\{\boldsymbol{f}_q^{x},...,\boldsymbol{f}_{q+N_2}^{x}\} \textbf{ s.t. } \theta_{er}^{2}\geq\{J(\boldsymbol{f}_q^{x}),...,J(\boldsymbol{f}_{q+N_2}^{x})\} \geq\{\theta_{0.5}^{2},\theta_{mcc}^{2}\}\\
&\vdots \\
S_{r}^{x}&=\{\boldsymbol{f}_q^{x},...,\boldsymbol{f}_{q+N_r}^{x}\} \textbf{ s.t. } \theta_{er}^{r}\geq\{J(\boldsymbol{f}_q^{x}),...,J(\boldsymbol{f}_{q+N_r}^{x})\} \geq\{\theta_{0.5}^{r},\theta_{mcc}^{r}\}
\end{IEEEeqnarray*}}\endgroup
where $\theta_{er}^{1}\geq\theta_{er}^{2}\geq\cdots\geq\theta_{er}^{r}$, ${\theta_{0.5}^{1}\leq\theta_{0.5}^{2}\leq\cdots\leq\theta_{0.5}^{r}}$, $\theta_{mcc}^{1}\leq\theta_{mcc}^{2}\leq\cdots\leq\theta_{mcc}^{r},$ $N_1\leq N_2\leq \cdots \leq N_r$, $q$ is an index that takes integer values in the range 1 to $L$ where $L=\gls*{numFilters}+2K+3$ and $J(.)$ is any of $\big[\text{E}_{r},\text{F}_{0.5},\text{MCC}\big]$, evaluated against the corresponding threshold. 
Each subset is used for training and testing the \ac{SVM} classifier by employing the cross-validation procedure described earlier. Their classification performance is evaluated using the \ac{AUC} of the \ac{ROC} curve \cite{Fawcett2004}. The subset that gives the highest \ac{AUC} is chosen.

The experimental framework is based on a systematic approach that aims to (a) find the best frame length $\gls*{genFrameSizems}$ for extracting the 5 alternative feature sets in (\ref{featspaces}), (b) examine the effect of the number of filters $\gls*{numFilters}$ on the classification performance when using either one of $\gls*{MelFreqSTFTfeatSet}$, $\gls*{LinFreqSTFTfeatSet}$, $\gls*{LfccfeatSet}$ or $\gls*{mfccfeatSet}$ and (c) obtain insights into the time-frequency information of normal and abnormal signals and their differences.
%
%
\section{Experiments, Results and Discussion}\label{sec:ChapSegmClasifExp}
An investigation of the effect of frame length values based on a deterministic approach was initially conducted by defining a suitable range and quantifying the classifier performance in order to choose the best $\gls*{genFrameSizems}$. Subsequent, Monte Carlo simulations were performed to test the suitability of the choice and to identify performance trends of the feature sets in a larger range. Finally, experiments were conducted varying the number of filters $\gls*{numFilters}$.

\subsection{Implementation details}
\label{implemDetails}
In all the experiments that follow, the signal segment length $\gls*{assumpWindow}$ was set to 20~s. Other time periods that do not violate the assumption outlined at the beginning of Section \ref{sec:featExtSection} could also be used but would affect the total number of segments obtained. Additionally, the error rate threshold $\gls*{errRateThres}$ was fixed at $0.456$ which is the error rate obtained when the predicted class is always the largest. This is the error rate attributed to random guessing and hence anything worse than this would mean that the classifier performs poorly. By keeping $\gls*{errRateThres}$ constant, the values of $\gls*{f05scoreThres}$ and $\gls*{mccThres}$ are varied in the range $[0,1]$ with step-size $w=0.05$ and the possible feature subsets are constructed and subsequently used in the classifier.

The L$_1$-norm soft margin formulation is used for \ac{SVM} due to its advantages over the L$_2$-norm in high dimensional feature spaces and in the presence of redundant features~\cite{Zhu2003a}. The sequential minimal optimization,~\cite{Rong-En2005}, is employed for solving the \ac{SVM} minimization problem which is the standard algorithm for this task. The penalty parameter, often called box constraint, is a term that trades off misclassification of training observations against simplicity of the decision surface. A low value makes the surface smooth (i.e. misclassification becomes less important), while a high value attempts to classify all training examples correctly. In the following experiments this parameter was set to 1. Finally, the formula used for the Gaussian kernel is $\exp(-\gamma \left\lVert \langle \boldsymbol{x}_1 - \boldsymbol{x}_2\rangle \right\rVert^2)$ for which $\gamma=1$ and $\langle\boldsymbol{x}_1,\boldsymbol{x}_2\rangle$ denotes the inner product between the training vectors ${x}_1$ and ${x}_2$.

\ac{LDA},~\cite{ONeil1992}, is also employed in certain experiments and the empirical pooled covariance matrix is used for the multivariate normal distribution of each class. Finally, for the \ac{CART} classifier, the split predictor (feature) is selected as the one that maximizes the split criterion gain (gini index) over all possible splits of all predictors~\cite{Loh2011a}. The tree, once fully grown, is pruned using the gini index as the pruning criterion.

\subsection{Data Acquisition and Test Protocol}
\label{dataAcq}
Adults with clinical knee \ac{OA} and reporting no knee pain in the last 2 weeks were recruited. Knees were classified by clinicians as: 1) normal (clinically healthy), or 2) abnormal (\ac{OA}). Exclusion criteria were: aged \textless18 years, previous surgery, unable to provide consent. \ac{AE1} signals were acquired during walking on a treadmill instrumented with force plates. The recordings were made with a sampling frequency of 44.1 or 48~kHz and downsampled to $\gls*{Fs}=16$~kHz for subsequent processing, 
using a contact microphone with a sound port for detecting airborne sounds and an electret condenser microphone mounted inside a capsule (Basik Pro Schertler, 20~Hz -- 20~kHz), 
attached over the patella and supported by a digital preamplifier (RME Babyface; PreSonus DigiMax LT). 

The assessment commenced with a 5 minute warm-up and acclimatisation to treadmill walking followed by data acquisition at increasing speeds on a flat level until maximum walking speed was achieved (speed increments of 0.5~km/h, maximum walking speed defined as the maximum pain-free speed where one foot was always in contact with the ground). Maximum speeds achieved ranged from 2.5 to 9~km/h. 


Data used in this work originates from 40 knees, of which 19 are normal (from 15 patients) and 21 are abnormal (from 18 patients).  
Table~\ref{table:subj_demogr} displays the demographical characteristics of the study participants. Approximately 83 minutes of sound data from healthy knees and 99 minutes from \ac{OA} is used. Following the segmentation process described in Section \ref{sec:featExtSection}, 249 normal and 297 abnormal segments of 20~s are obtained.

\begin{table}[!t]
	\centering
	\caption{Demographical characteristics of study participants. Age and BMI are reported as mean $\pm$ standard deviation. Numbers inside brackets denote minimum and maximum values. \label{table:subj_demogr}}
	\begin{tabular}{>{\centering\arraybackslash}*3c}
		\toprule
		{}   & Healthy   & OA \\
		\midrule
		\multicolumn{1}{l}{Participants}   & 19              &  21             \\
		\multicolumn{1}{l}{Females/males}  & 5/14            &  11/10            \\
		\multicolumn{1}{l}{Age (years)}    & $40.1 \pm 18.3~[21.3, 80.0]$ &  $62.6 \pm 14.4~[28.7, 80.4]$\\
		\multicolumn{1}{l}{BMI (kg/m$^2$)} & $23.7 \pm 2.9~[19.2, 28.5]$  &  $29.0 \pm 6.4~[21.0, 42.2]$ \\
		\bottomrule
	\end{tabular}
\end{table}

\subsection{Deterministic search in a specified range of frame lengths}
\label{expdeterm}
%
The frame length $\gls*{genFrameSizems}$ is tested for the values $\big[20,24,28,\ldots,100\big]$ ms. The limiting values were chosen so that 20~ms is a short enough window to allow good time resolution in the time-frequency representations for the sounds (clicks, pops, grindings) heard during walking and 100~ms is a large enough window to capture the two major events in a single stride, namely, the heel strike and the push off responses as captured by the patella microphone. This information was extracted from the ground reaction force signals obtained from the treadmill's force plates from which can be extracted the timings of each event in the gait cycle.

The experimental framework developed in the previous Sections is applied and, for each frame length $\gls*{genFrameSizems}$, the features are extracted, analysed independently and the possible subsets are constructed for evaluation using the $\text{SVM}_{l}$ classifier. In all cases $\gls*{numFilters}=20$ is chosen, giving 60 coefficients (including the 0\textsuperscript{th} cepstral coefficient) for all the feature sets except $\gls*{STFTfeatSet}$. For $\gls*{STFTfeatSet}$, the size depends on $\gls*{genFrameSizems}$ since the \ac{DFT} length used is equal to the frame length, as shown in Section \ref{sec:featExtSection}.

\begin{figure}[t]
	\centering
	\includegraphics{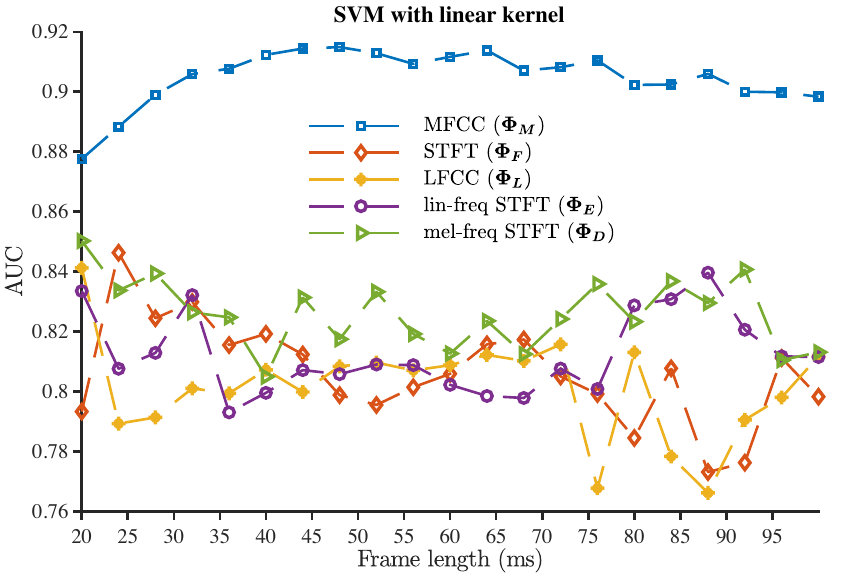}
	\caption{AUC against frame length for SVM (linear kernel). The points are connected with dashed lines to aid the visualisation. \label{fig:aucSVM_lin}}
\end{figure}

For the analysis of the results, a top-down approach is followed. The overall results are summarised in Fig.~\ref{fig:aucSVM_lin}. Each point on a line represents the highest \ac{AUC}, averaged over 100 trials, obtained by any subset of the corresponding frame-size and of the particular feature set.

It is evident from  Fig.~\ref{fig:aucSVM_lin} that classifying using features obtained from the $\gls*{mfccfeatSet}$ feature set, including first and second derivatives, scores consistently higher than any other set, for any $\gls*{genFrameSizems}$, with its best performance occurring with $\gls*{genFrameSizems}=$ 48~ms $(\text{AUC}=0.915)$. On the other hand, the sets $\gls*{STFTfeatSet}$, $\gls*{LfccfeatSet}$, $\gls*{LinFreqSTFTfeatSet}$ and $\gls*{MelFreqSTFTfeatSet}$ peak at 24~ms, 20~ms, 88~ms and 20~ms respectively. Initial observations suggest that the optimal frame-size is different for each feature set and that reducing the dimensionality of the spectrum with triangular shaped filters generates comparable results with the full spectrum features and in some cases ($\gls*{genFrameSizems}\geq$ 72~ms) improves the classification performance. 
%

\begin{figure}[t]
	\centering 
	\includegraphics{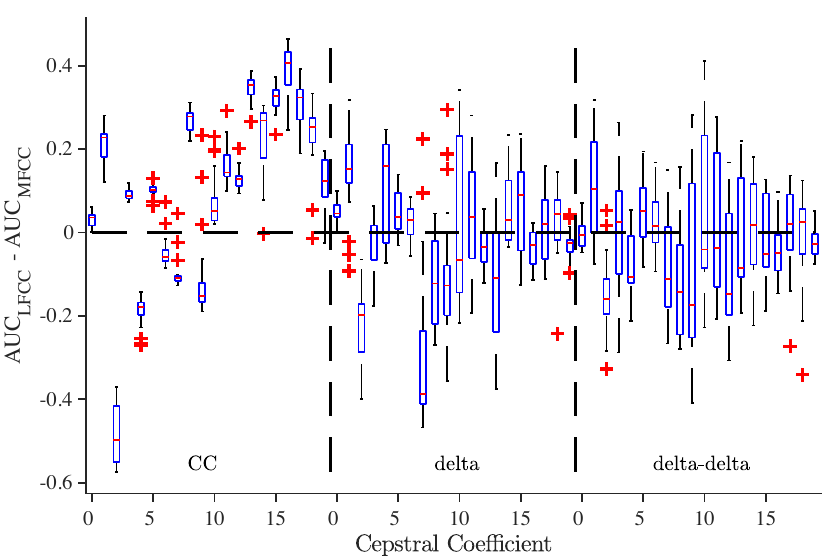} 
	\caption{Difference in AUC values of LFCC and MFCC for all frame lengths\label{fig:lfmf_auc}.}
\end{figure}
\begin{figure*}[t]
	\centering
	\includegraphics{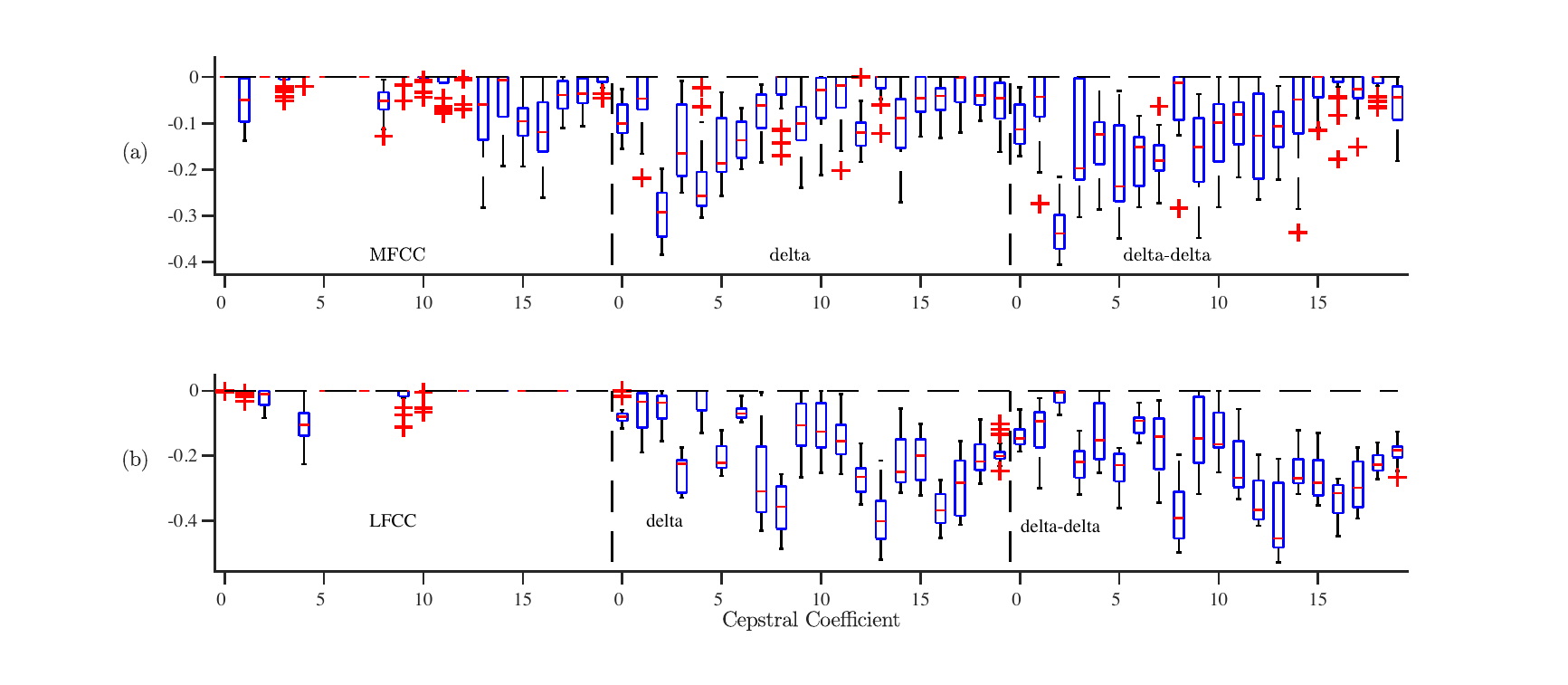} 
	\caption{Comparison of AUC output for the static, delta and delta-delta cepstral coefficients for MFCC (a) and LFCC (b), in all frame lengths.\label{fig:ccvsdelta_auc}}
\end{figure*}
\subsubsection{Comparison of MFCC and LFCC results}
It is seen in Fig. \ref{fig:aucSVM_lin} that \ac{AUC} values range from $0.878$ to $0.915$ and from $0.77$ to $0.84$ for $\gls*{mfccfeatSet}$ and $\gls*{LfccfeatSet}$ respectively. Hence, the performance for $\gls*{mfccfeatSet}$ is only weakly sensitive to the choice of frame-size and in fact, above 32~ms the variance of the metric value drops approximately to $\nicefrac{1}{3}$ of the variance obtained when including the \ac{AUC} for $l<32$~ms. The fluctuation is small because, for any $\gls*{genFrameSizems}$, the final selected subset that gave the highest \ac{AUC} consisted of only $\boldsymbol{f}^M_3$ which contains the statistical parameters of the distribution of the $2^{\text{nd}}$ \ac{MFCC}. All other subsets considered from this set resulted in worse performances. For $\gls*{genFrameSizems}\geq88$~ms and $\gls*{genFrameSizems}\leq32$~ms, the \ac{AUC} of $\gls*{mfccfeatSet}$ follows a downward trend suggesting that, even to a small degree, the performance drops for frame-sizes outside of this range. This hypothesis is tested in Section \ref{ssec:mcsimuls}. Higher variability is observed in the results of $\gls*{LfccfeatSet}$ where more than one feature vector is chosen in the final subset for $10$ out of $21$ frame lengths. The highest \ac{AUC} for this set occurs at $20$~ms which is obtained with a subset of $17$ feature vectors. The classification rate for this case, averaged over the $100$ cross-validation iterations, is $75.07\%$ compared to $85.25\%$ obtained from the $\gls*{mfccfeatSet}$ feature set with $\gls*{genFrameSizems}=48$~ms. This translates into classifying correctly, $55$ more segments (from a total of $546$ segments) when using the $\gls*{mfccfeatSet}$ set.

Fig.~\ref{fig:lfmf_auc} shows a box plot of the differences in \ac{AUC} for all frame lengths per \acf{CC}. It is evident that many \ac{CC} obtained with a linear-frequency filter-bank generated better classification results than their mel-frequency filter-bank counterparts. More precisely, the feature vectors corresponding to high order $(10$ to $19)$ as well as low order coefficients $(0,1,3,5,8)$ consistently give higher \ac{AUC} values in any of the frame length cases tested. Exceptions occur in each of the $14^{\text{th}}$ and $18^{\text{th}}$ coefficient at 20~ms and one in the $19^{\text{th}}$ at 28~ms but can be considered as outliers because of their very small value. The feature vector representing the $2^{\text{nd}}$ \ac{MFCC}, which was found to be the most important of the $\gls*{mfccfeatSet}$ set, clearly outperforms the $2^{\text{nd}}$ \ac{LFCC} with a difference that reaches $-0.57$ in the \ac{AUC} result (absolute median value is $0.5$). The same holds for the $4^{\text{th}}$ and $9^{\text{th}}$ \ac{CC} but with a smaller absolute median value of $0.18$ and $0.15$ respectively. For the trajectory coefficients however, there is no apparent advantage of one feature set over the other that persists with changing frame length, except in very few cases (at the $2^{\text{nd}}$ and $7^{\text{th}}$ delta coefficients).

Fig. \ref{fig:ccvsdelta_auc} compares the static coefficients with their time derivatives for \ac{MFCC} in (a) and \ac{LFCC} in (b). 
The top performing coefficient is found, per feature set and for each index $(0$ to $19)$, by comparing the \ac{AUC} scores of the static coefficient and its delta and delta-delta. Each box in the figure consists of the values obtained, for every frame length, by subtracting the \ac{AUC} of the coefficient indicated by the index on the x-axis from the top performing one found as before. 
The \ac{MFCC} static coefficients $0, {2\text{ to } 7}\text{ and } 9\text{ to } 12$ exhibit better classification performance in over 65\% of the frame lengths with the $0^{\text{th}}$, $2^{\text{nd}}$ and $5^{\text{th}}\text{ to } 7^{\text{th}}$ scoring consistently higher than their trajectories in any frame length. This is different for the $1^{\text{st}}$, $8^{\text{th}}$ and $13^{\text{th}}\text{ to } 18^{\text{th}}$ static \acp{CC} where they score lower than either the delta or delta-delta coefficients in over 50\% of the frame lengths. The feature vector representing the $19^{\text{th}}$ \ac{MFCC} scores similarly with the vectors corresponding to the coefficient's derivatives but its performance degrades for $\gls*{genFrameSizems}>84$~ms. For the $\gls*{LfccfeatSet}$ feature set, the overwhelming majority of the static \acp{CC} score consistently higher than their delta and delta-delta. More specifically, the feature vectors $\boldsymbol{f}_i^L$ for $i=4,\ldots,20, i \neq5,10,11$ score higher for any $\gls*{genFrameSizems}$. 

Overall, the results show that the higher order \ac{LFCC} outperform the corresponding \ac{MFCC} (Fig.~\ref{fig:lfmf_auc}) but Fig.~\ref{fig:aucSVM_lin} suggests that, from a classification point of view, it is better to use mel-scaled instead of linear-scaled filters to extract the \ac{CC}. However, to justify this, further tests are needed in order to investigate the effect on the performance, the number of filters in the filter-bank has. This is explored in Section \ref{ssec:nbandssection}.

\subsubsection{Comparison of STFT and compressed STFT feature sets}
\begin{figure}[t]
	\centering
	\includegraphics[width=8.5cm,keepaspectratio]{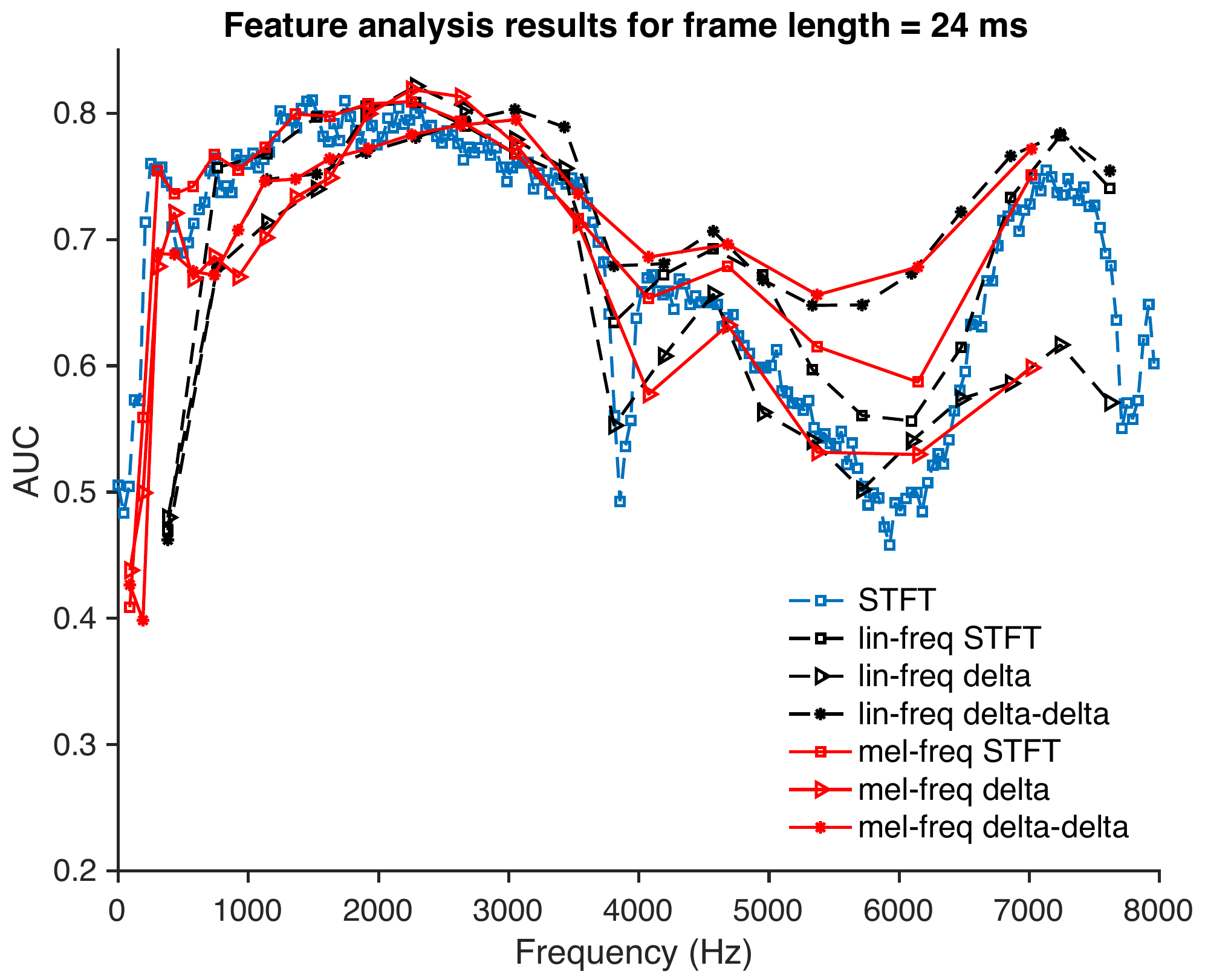} 
	\caption{Individual performance of STFT based feature vectors at $l=24$~ms with $N_B=20$.\label{fig:auc24stft}}
\end{figure}

All three feature sets, $\gls*{STFTfeatSet}$, $\gls*{LinFreqSTFTfeatSet}$ and $\gls*{MelFreqSTFTfeatSet}$, achieve comparable maximum performances in \ac{AUC}, with values 0.846 $(\gls*{genFrameSizems}=24$~ms$)$, $0.839$
$(\gls*{genFrameSizems}=88$~ms$)$ and $0.850$ $(\gls*{genFrameSizems}=20$~ms$)$ respectively. 
At $\gls*{genFrameSizems}=24$~ms the subset selected from $\gls*{STFTfeatSet}$ that gave the highest \ac{AUC} only consists of features that fall within the range $0.29$ to $2$~kHz. Comparing the performances of the individual feature vectors (i.e. the 11-dimensional elements of $\gls*{LinFreqSTFTfeatVec}$, $\gls*{MelFreqSTFTfeatVec}$ and $\gls*{STFTfeatVec}$) within this range, shows that the $\gls*{STFTfeatVec}$ feature vectors perform better than the corresponding from $\gls*{LinFreqSTFTfeatVec}$ and $\gls*{MelFreqSTFTfeatVec}$. This is depicted in Fig.~\ref{fig:auc24stft} that shows the average \ac{AUC} scores per feature vector, against frequency. 
At approximately $3.5$~kHz however, the performance starts to drop rapidly, becoming comparable to, or worse than, the corresponding features from the $\gls*{LinFreqSTFTfeatSet}$ and $\gls*{MelFreqSTFTfeatSet}$ sets. At around $6$~kHz the performance starts to improve again, reaching a maximum near 7.2~kHz at a value of $0.75$, comparable to that of the lower frequency features ($\leq3.5$~kHz). Above $4.5$~kHz it is observed that the classification performance actually improves when using the static coefficients or their second derivatives from the $\gls*{LinFreqSTFTfeatSet}$ and $\gls*{MelFreqSTFTfeatSet}$ sets. 
The line plots in Fig.~\ref{fig:auc24stft} follow similar trends.

%

\begin{figure}[t]
	\centering
	\includegraphics[width=8.5cm,keepaspectratio]{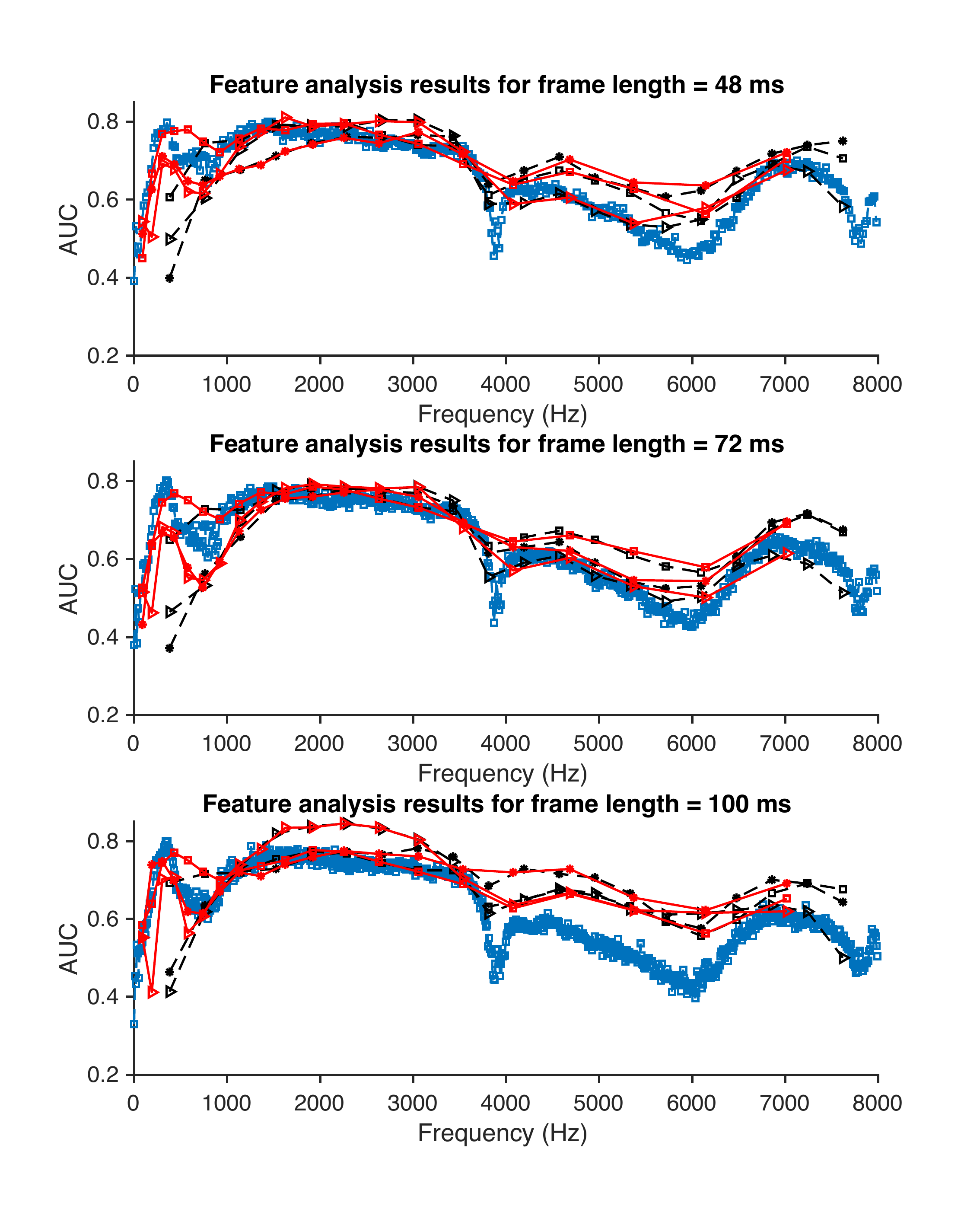} 
	\caption{Individual performance of STFT based feature sets at (a) $l=48$~ms, (b) $l=72$~ms and (c) $l=100$~ms with $N_B=20$. Line styles and colours are the same as in Fig. \ref{fig:auc24stft} \label{fig:AUC_stft_examples}.}
\end{figure}

Adding to the above, the static coefficients from both $\gls*{LinFreqSTFTfeatSet} $ and $\gls*{MelFreqSTFTfeatSet}$ perform better than their corresponding derivatives for frequencies up to $2$~kHz. The delta coefficients peak in the range $2.2$ to $2.6$~kHz but then begin to drop rapidly whereas for higher frequencies $(\geq5$~kHz$)$ the delta-delta coefficients perform better overall. This shows that additional information exists in the dynamics of the spectrum for mid to high frequencies. The above observations can also be derived from the results at other, higher values of $\gls*{genFrameSizems}$. From the plots of Fig.~\ref{fig:AUC_stft_examples} it becomes apparent that, with increasing frame length, the delta coefficients capture more information than the corresponding static and delta-delta coefficients in the frequency range $1.6$ to $3$~kHz.

With increasing frame length, the frequency resolution improves and it becomes immediately apparent from Fig.~\ref{fig:AUC_stft_examples} that two regions of high performing features exist, one in the range $[220,420]$~Hz, denoted as $F_{r_1}$, and another in the range $[1,3.4]$~kHz, denoted as $F_{r_2}$. These frequency regions show a collection of features that individually score higher than the rest in their respective set (e.g. ${\geq0.726}$ \ac{AUC} in the $\gls*{STFTfeatSet}$ set). It is also worth noticing that for $\gls*{genFrameSizems}\leq36$~ms there exists another frequency band $(6.6$ to $7.6$~kHz$)$, where some feature vectors that fall within the band, from all three sets $(\gls*{STFTfeatSet}$, $\gls*{LinFreqSTFTfeatSet}$, $\gls*{MelFreqSTFTfeatSet})$, also scored high \ac{AUC} $(\geq 0.75)$. However, in the three examples of Fig.~\ref{fig:AUC_stft_examples} it is seen that their performance gradually drops with increasing frame length to, or below, $\text{AUC}=0.5$. This means that the classifier randomly assigns observations to classes $(\text{AUC}=0.5)$, given the input features, or the classifier failed to apply the information at hand correctly $(\text{AUC}<0.5)$. For the second case one may reverse the classifier's decisions and obtain a \ac{ROC} curve that would give $\text{AUC}>0.5$, as long as the classifier consistently produces results falling in the lower right part of the \ac{ROC} space as described in~\cite{Fawcett2006}.

Feature vectors from $\gls*{STFTfeatSet}$ that fall within $F_{r_1}$ consistently perform better than the corresponding vectors from both $\gls*{MelFreqSTFTfeatSet}$ and $\gls*{LinFreqSTFTfeatSet}$ in all frame-sizes tested. 
On the contrary, the performance of the feature vectors from the latter two sets, at frequencies that fall within $F_{r_2}$, increases with increasing $\gls*{genFrameSizems}$ and becomes similar to or even better than those of the $\gls*{STFTfeatSet}$ set. As can be seen in Fig.~\ref{fig:AUC_stft_examples}~(c), for frequencies between 1.5~kHz and 3~kHz (within $F_{r_2}$), the delta coefficients from both $\gls*{LinFreqSTFTfeatSet}$ and $\gls*{MelFreqSTFTfeatSet}$ sets outperform the rest, achieving \ac{AUC} $\geq0.8$. This effect is reflected in the performance curve (Fig. \ref{fig:aucSVM_lin}) of the $\gls*{LinFreqSTFTfeatSet}$ and $\gls*{MelFreqSTFTfeatSet}$ feature sets as a slightly upward trend compared to the degrading performance with the $\gls*{STFTfeatSet}$ set.  

The feature subsets from the $\gls*{STFTfeatSet}$ set, selected based on the method described in Section \ref{sec:analAndSelectSection}, contain only features that fall within $F_{r_1}$ or $F_{r_2}$ for $l\geq24$ and only from $F_{r_1}$ for $\gls*{genFrameSizems}\geq32$. For the $\gls*{MelFreqSTFTfeatSet}$ and $\gls*{LinFreqSTFTfeatSet}$ sets, the majority of the feature vectors included in the selected subsets fall within the two frequency bands for any $\gls*{genFrameSizems}$ tested. Therefore, these observations highlight the importance of the spectrum features that fall within the two identified bands of frequencies and show that these features have a strong impact on the classification performance.

In the analysis of our previous work, \cite{Yiallourides2018}, it was observed that the top performing features obtained from the magnitude spectrum are primarily at the low frequencies. This is further supported here where we have also identified specific bands that carry significantly discriminant information.

From the above results and observations it can be concluded that the information in the time-frequency spectrum that enables an \ac{SVM} classifier to discriminate between the two classes (normal/healthy vs abnormal/\ac{OA}) is contained in a range of frequencies but in specific bands that depend on $\gls*{genFrameSizems}$. It can be deduced that the frame-size, and hence the \ac{DFT} length, is a classification performance trade-off. Small $\gls*{genFrameSizems}$ is preferred in order to achieve suitable frequency resolution for the knee sounds occurring in the range of $0.7$ to $3.5$~kHz and at frequencies $\geq6$~kHz but a larger $l$ is preferred to capture the finer details of the spectrum in the $220-420$~Hz band. For the sets $\gls*{MelFreqSTFTfeatSet}$ and $\gls*{LinFreqSTFTfeatSet}$, the latter two hold true. However, for the range $0.7$ to $3.5$~kHz the choice of $\gls*{genFrameSizems}$ does not significantly affect the maximum \ac{AUC} values achieved but it has an effect in the performance of the individual coefficients. 

The suitability of the choice $\gls*{numFilters}=20$ is tested in Section \ref{ssec:nbandssection}, even though reducing the dimensionality of the \ac{STFT} spectrum using 20 filters scaled either linearly or non-linearly (mel) in frequency improves the \ac{AUC} in 20 out of the 21 $\gls*{genFrameSizems}$ values tested. An adaptable filter-bank with narrower filters at the low $(<1$~kHz$)$ and highest frequencies $(\geq6$~kHz$)$ and broader at the mid to high frequencies could capture, with fewer coefficients than the full resolution spectrum, the information needed to discriminate the normal and abnormal signals with higher than 0.85 in \ac{AUC}. This is further supported by the results showing that the $\gls*{MelFreqSTFTfeatSet}$ low frequency features $($up to 500~Hz or 1.6~kHz depending on $\gls*{genFrameSizems})$ consistently perform better than the corresponding $\gls*{LinFreqSTFTfeatSet}$ whereas at high frequencies the difference in \ac{AUC} is diminished and sometimes reversed. These observations are further investigated and validated in Section~\ref{ssec:nbandssection}.

\subsection{Local search in the vicinity of the best frame length}
The $\gls*{genFrameSizems}$ values that gave the maximum \ac{AUC}, denoted as $l_{b,x}$ for the frame length $b$ of the $x$ feature set, were found from the experiments in the previous Section in which the time step used was $t=4$~ms. The existence of local maxima in the vicinity of these points can be tested by defining a grid of 6 values with a time step of 1~ms centred at $l_{b,x}$ i.e. $l_{b,x}-t$ for $t=3,2,\ldots,-3, t\neq0$. Steps shorter than 1~ms will not have an impact in the result given that, at $\gls*{Fs}=16$~kHz, the sample difference would be less than 16 samples.

The same experimental framework is followed and the training and test sets are standardized as before. 
In this experiment, 3 more classifiers are used to evaluate the subsets created at the feature selection step in order to assess how well the subsets generalise with different classifiers. 
These are the (a) \ac{LDA} classifier that finds linear hyperplanes in the feature space which separate the two classes, (b) \ac{CART} and (c) \ac{SVM} with a Gaussian kernel ($\text{SVM}_{g}$) in order to look for more complex and non-linear boundaries in the feature space. Classification results are again evaluated using \ac{AUC}.

From these experiments it was found 
that, for the sets $\gls*{mfccfeatSet}$ and $\gls*{LfccfeatSet}$, $\text{SVM}_{l}$ and \ac{LDA} generated comparable results at approximately \ac{AUC}$=0.9$ and $0.8$ in the respective sets, with $\text{SVM}_{l}$ being slightly better by $1\%$ for $\gls*{mfccfeatSet}$ and $0.5\%$ to $7\%$ for $\gls*{LfccfeatSet}$. $\text{SVM}_{g}$ and \ac{CART} achieved \ac{AUC} scores that ranged from $9\%$ to $55\%$ lower than the maximum. 
For both $\gls*{LinFreqSTFTfeatSet}$ and $\gls*{MelFreqSTFTfeatSet}$ sets, $\text{SVM}_{l}$ outperformed by at least $5.2\%$ and $10.5\%$ respectively the other classifiers which generated comparable scores. For all ${l_{b,x}-t}$ in $\gls*{STFTfeatSet}$ and $4$ out of the $6$ values of $t$ in $\gls*{LfccfeatSet}$, $\text{SVM}_{g}$ performed significantly worse than the other classifiers by at least $70\%$ and $31\%$ respectively, producing results as low as $\text{AUC}=0.36$. The power of the \ac{SVM} classifier with a Gaussian kernel is limited in this case by the relatively large number of features compared to the small training set size, which increases the risk of overfitting when the data is transformed to a high dimensional feature space. Therefore, for the binary classification task of this work and based on the amount of available data, the performance advantage of linear classifiers suggests that the two classes can be linearly separated in the feature spaces explored.
\begin{table}[t]
	\centering
	\caption{Average cross-validation results of the best subsets per feature set using linear kernel SVM.\label{table:svmlinAvgRes}}
	\begin{tabular}{>{\centering\arraybackslash}P{0.2cm} c c c c c P{0.4cm} P{1.3cm}}
		\toprule
		{\scriptsize Feature set} & \multirow{2}{*}{AUC} & \multirow{2}{*}{$\gls*{errRate}$} & \multirow{2}{*}{$\gls*{f05score}$} & \multirow{2}{*}{$\text{MCC}$} & \multirow{2}{*}{$\gls*{s_score}$} & $\gls*{genFrameSizems}$ (ms) & {\scriptsize Feature vectors used}\\
		\midrule
		\multicolumn{1}{c}{$\gls*{mfccfeatSet}$}                  & 0.917  & 0.147 & 0.853 & 0.705 & 0.804  & 49  & 1 \\ 
		\multicolumn{1}{c}{$\gls*{LfccfeatSet}$}                  & 0.841  & 0.249 & 0.723 & 0.501 & 0.658  & 20  & 17 \\ 
		\multicolumn{1}{c}{$\gls*{STFTfeatSet}$}  & 0.848  & 0.218 & 0.756 & 0.564 & 0.701  & 23  & 11 \\ 
		\multicolumn{1}{c}{$\gls*{LinFreqSTFTfeatSet}$}  &0.844  &0.239  &0.723 &0.536 &0.673  &90  &2  \\ 
		\multicolumn{1}{c}{$\gls*{MelFreqSTFTfeatSet}$}  &0.875  &0.195  &0.780 &0.611 &0.732  &21  &20  \\ 
		\bottomrule  
	\end{tabular}
\end{table}

As described in the previous paragraph, $\text{SVM}_{l}$ achieves higher \ac{AUC} with all of the feature sets. However, the features were selected based on results obtained from training and testing this specific type of classifier. Therefore, the subsets created at the feature selection step are tailored to work better with this and similar classifiers. Nevertheless, the work in this paper is not concerned with finding the best classifier to use with the available data. It is rather focused on the efficacy of the features in question to separate the two classes and capture information that would eventually lead to finding the specific abnormality (\ac{OA}) signatures in the signals.

The experiments showed that the average \ac{AUC} values for the frame-sizes $l_{b,x}-t$, for $t=3,2,\ldots,-3, t\neq0$, were close to those of $l_{b,x}$. 
%
The results are improved for all feature sets except for $\gls*{LfccfeatSet}$. Table \ref{table:svmlinAvgRes} reports the frame length value that gave the highest \ac{AUC} per feature set, found using $\text{SVM}_{l}$. A corresponding score, $\gls*{s_score}$, computed as ${\gls*{s_score}=\big[\text{MCC}+(1-\text{E}_r)+\text{F}_{0.5}\big]/3}$,~\cite{Yiallourides2018}, 
is also used and consists of the \ac{MCC} and $\gls*{f05score}$ measures which capture different attributes of the classification result than the \ac{AUC} and would therefore be useful in the analysis. 
$\gls*{s_score}$ can vary between 0 and 1 (where $\gls*{s_score}=1$ indicates perfect prediction). It is shown in Table~\ref{table:svmlinAvgRes} that $\gls*{s_score}$ ranks the top performance of the feature sets in the same way as \ac{AUC}.

\subsection{Monte Carlo experiments}
\label{ssec:mcsimuls}
In the previous experiments, the $\gls*{genFrameSizems}$ values to be tested were defined in a deterministic approach. In this Section, a stochastic approach is followed in which the values are randomly chosen from a range. In this way the optimality of the result in the previous Section is assessed and the effect of increasing or decreasing the frame length even further is examined.

Firstly, plausible limits for $\gls*{genFrameSizems}$ need to be set. Using frames larger than a single stride will cause overlap of the sounds from two strides resulting in poor modelling of those sounds. In addition, applying the \ac{DFT} operation will become inappropriate because the signal in a single frame will be non-stationary. 
%
The maximum recorded speed is $9$~km/h giving an average stride duration of $0.7$~s. Given that a single frame length value is applied at the feature extraction stage for all $\gls*{iKneeSigjSegm}$, the upper bound is set at $0.7$~s. For the lower bound $2$~ms is chosen which gives a time resolution that allows fine localisation of the sounds and approximately $470$~Hz frequency resolution at $\gls*{Fs}=16$~kHz for the \ac{DFT}.
\begin{figure}[t]
	\centering
	\includegraphics{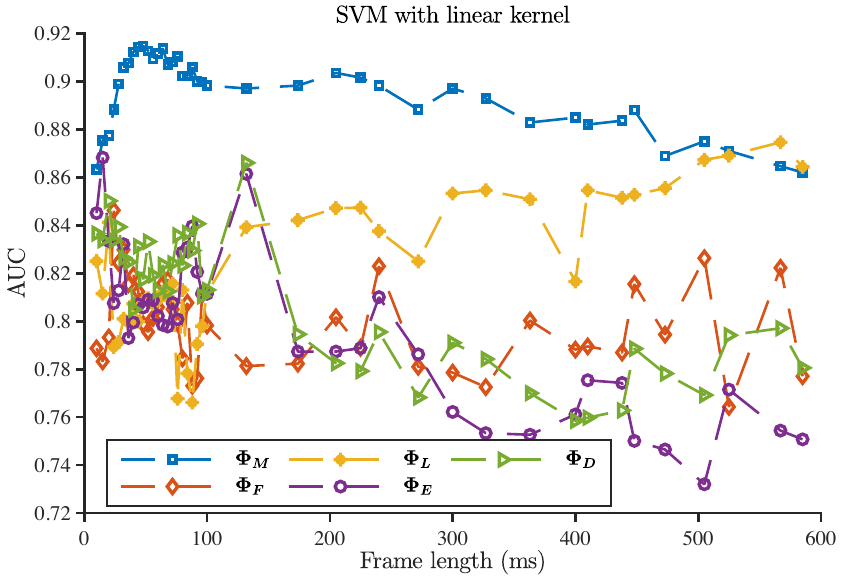} 
	\caption{AUC per feature set against frame length for SVM (linear kernel) - including the Monte Carlo results\label{fig:mcsimulplot}.}
\end{figure}
%
\begin{figure*}[t]
	\centering
	\includegraphics[width=\textwidth,keepaspectratio]{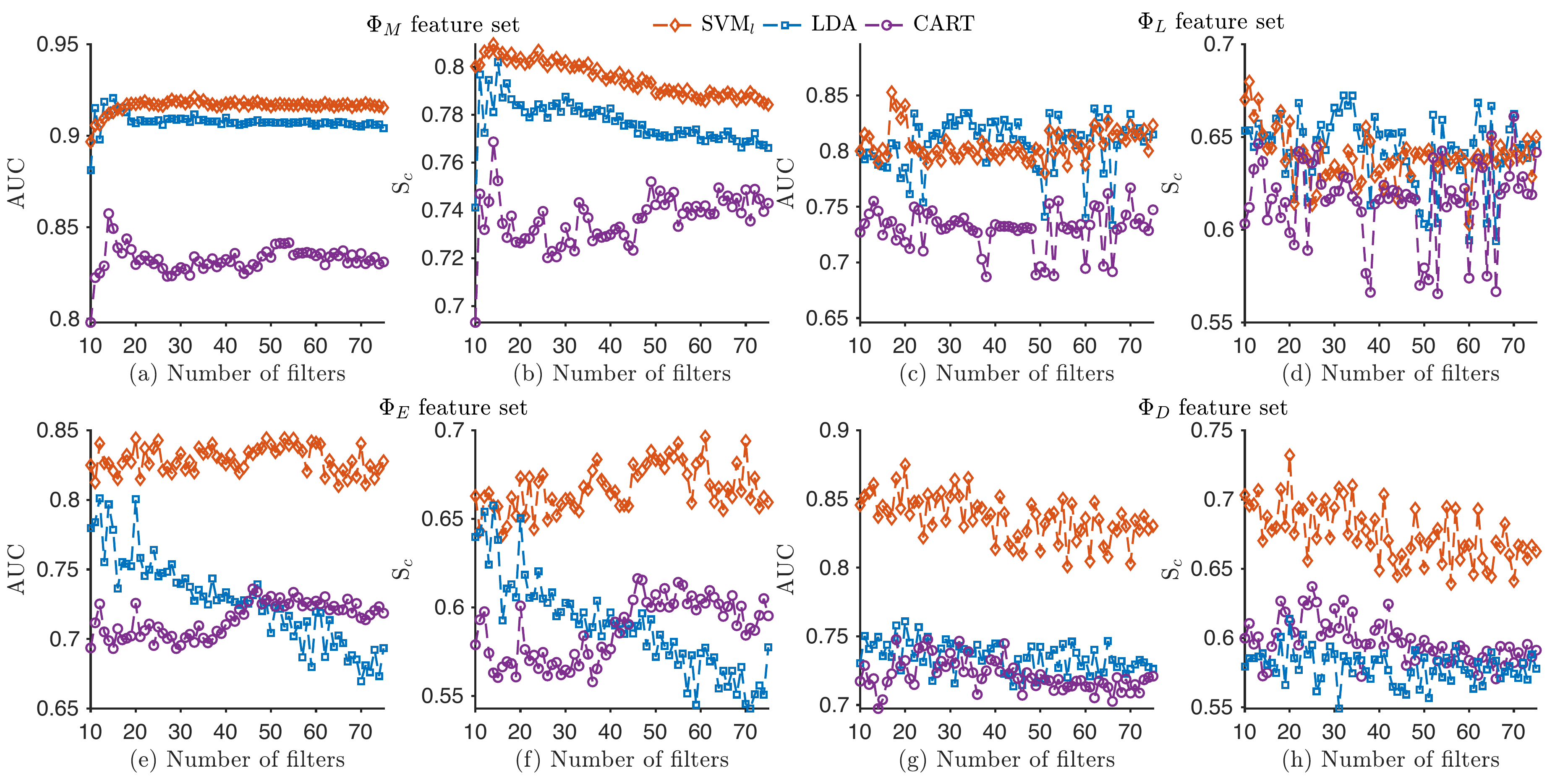} 
	\caption{Effect of the number of filters, $N_B$, on classification performance per feature set, using 3 different classifiers.\label{fig:nbandsRes_2}}
\end{figure*}

Monte Carlo simulations are performed by randomly assigning $20$ values to $\gls*{genFrameSizems}$ within the range $2$ to $700$~ms, excluding the range $20$ to $100$~ms, since it was examined in the previous experiments. The same experimental framework is executed as before using only $\text{SVM}_{l}$ and the results are shown in Fig.~\ref{fig:mcsimulplot}. 
%
In comparison to Table \ref{table:svmlinAvgRes}, an improvement of $2.8\%$ is observed for the $\gls*{LinFreqSTFTfeatSet}$ set at $15$~ms, giving an \ac{AUC} of $0.868$ compared to the previous $0.844$. 
With this experimental framework and using fixed frame segmentation, only the $\gls*{mfccfeatSet}$ feature set generates a classification performance 
that behaves smoothly over the entire range and has a global maximum. 
%
The $\gls*{LfccfeatSet}$ feature set on the other hand generates a clear local maximum in the range of $20$ to $100$~ms. However, for $\gls*{genFrameSizems}\geq205$~ms (except at $272$ and $400$~ms) the \ac{AUC} scores are higher than the previously best one obtained at $\gls*{genFrameSizems}=20$~ms. The maximum is achieved at $\gls*{genFrameSizems}=567$~ms giving $\text{AUC}=0.875$ which is even higher than that achieved with $\gls*{mfccfeatSet}$. The $\gls*{MelFreqSTFTfeatSet}$ and $\gls*{LinFreqSTFTfeatSet}$ feature sets achieve an \ac{AUC} of $0.86$, both at $132$~ms, whereas for $\gls*{genFrameSizems}\geq174$~ms, \ac{AUC} values less than $0.81$ are achieved with both sets. The $\gls*{STFTfeatSet}$ set on the other hand exhibits a more variable behaviour, with the \ac{AUC} ranging between $0.76$ and $0.85$. However, it shows a clear global maximum at $\gls*{genFrameSizems}=23$~ms.
%
\subsection{Experiments on the number of filters}
\label{ssec:nbandssection}
The effect of the $\gls*{numFilters}$ parameter on the classification performance of all the feature sets except $\gls*{STFTfeatSet}$ is examined in this section. 
$\gls*{numFilters}$ is varied from $10$ to $75$ while $l$ is kept fixed at the values of Table \ref{table:svmlinAvgRes}. 
%
%
The outcome is evaluated using \ac{AUC} and $\gls*{s_score}$. The experimental outcomes are summarised in Fig.~\ref{fig:nbandsRes_2} for each set, showing the highest values obtained by any feature subset in each $\gls*{numFilters}$ case based on the plots' metric. As expected, the specific values of the two performance metrics are classifier depended. However, the general outcomes and the observations derived are similar across the different classifiers.

Fig. \ref{fig:nbandsRes_2} (a) and (b) show an overall negative trend with a small variance in the final value in both metrics as $\gls*{numFilters}$ increases. It can be deduced that there is a strong indication that a small number of filters $(\leq20)$ is more suitable for the extraction of $\gls*{MelDftMat}$. The two metrics are shown to improve slightly for ${\gls*{numFilters}>50}$ only with \ac{CART} but later drop and never exceed the highest score obtained with $\gls*{numFilters}=14$. With $\text{SVM}_l$ and \ac{CART}, the subsets that generated the highest results for any $\gls*{numFilters}$ always included $\boldsymbol{f}^M_3$. In fact, $\forall \gls*{numFilters} \neq 10,11,13$, this particular feature vector was the only one selected in the final subset, one of which generated the overall highest result $(\text{AUC}=0.921$ for $\gls*{numFilters}=33$ and $\text{SVM}_l)$. For \ac{LDA}, the selected feature subsets included $\boldsymbol{f}^M_3$ together with at most 2 more features (both static and delta coefficients) for $\gls*{numFilters}=11,12,\ldots,18$ whereas for all other cases $\boldsymbol{f}^M_3$ was the only feature vector in the subset.

The maximum classification performance measured in both \ac{AUC} and $\gls*{s_score}$ is achieved when using only $\boldsymbol{f}^M_3$. However, if an exhaustive search is performed through the entire $\gls*{mfccfeatSet}$ feature set and all possible combinations are used with a classifier, it is likely that a subset containing more vectors than only $\boldsymbol{f}^M_3$ would generate better performance. For this experiment however, such a method is computationally very costly as it would generate $\sum_{n=1}^{3\gls*{numFilters}} {3\gls*{numFilters}\choose n}$ possible combinations per classifier. Given that there is only a limited amount of data, the classifier results would become meaningless when $n$ becomes larger than a certain value because the feature space will eventually become very sparse. When this happens, the classifier's decision boundaries will be formed due to the sparseness of the feature space and not due to the information captured by the features which can also lead to overfitting, among other problems,~\cite{Theodoridis2009}. For these reasons, a suboptimal search method, like the one employed in this work, is favoured and was found to generate good performance as shown in Fig.~\ref{fig:nbandsRes_2} (a) and (b). 


Compared to $\gls*{mfccfeatSet}$, the results of the $\gls*{LfccfeatSet}$ feature set are more variable as shown by plots (a) to (d) in Fig.~\ref{fig:nbandsRes_2}. High classification scores are obtained for $13\leq \gls*{numFilters} \leq 50$ (depending on the classifier) and the best result for this set is obtained with $\text{SVM}_l$ and $\gls*{numFilters}=17$ giving an \ac{AUC} of 0.853. A slightly upward trend for $\gls*{numFilters}\geq51$ is observed, which is more noticeable with $\text{SVM}_l$. 
%
%
From Fig.~\ref{fig:nbandsRes_2}~(e) and (f), it is clear that, when using \ac{LDA} with the $\gls*{LinFreqSTFTfeatSet}$ set, the classification performance in terms of \ac{AUC} and $\gls*{s_score}$ is higher for $\gls*{numFilters}\leq20$. With \ac{CART}, an increase in the values of both metrics is observed for $\gls*{numFilters}\geq46$ which exceeds that achieved with \ac{LDA} for those values. This jump is attributed to the inclusion of features falling in the frequency band $F_{r_1}$ compared to only using features that fall within $F_{r_2}$ for $\gls*{numFilters}<32$. This effect is less obvious with $\text{SVM}_l$ where the scores achieved in the range of the $\gls*{numFilters}$ values tested are comparable. For the final set, $\gls*{MelFreqSTFTfeatSet}$, the plots are similar amongst the classifiers and suggest that, for best results, a small number of filters $(\gls*{numFilters}\leq20)$ is also preferable. It is observed that the selected subsets with which the classifiers scored the highest \ac{AUC} and $\gls*{s_score}$ only contain feature vectors that fall in either $F_{r_1}$ or $F_{r_2}$ or both. This further supports the observations and results discussed in Section \ref{expdeterm}, stressing the importance of these frequency bands, as well as the importance of the information in the dynamics of the spectrum for classifying normal and abnormal knee signals.

\section{Conclusions}
\label{conclusions}

This paper 
%
investigates 
the discriminant power of time-frequency features for the task of classifying knee condition, and explores effective parameterisations of the knee sounds collected from healthy and \ac{OA} knees during walking. The efficacy of knee condition classification was evaluated by qualitative and quantitative analysis using the \ac{AUC} of the \ac{ROC} curve and metrics derived from the confusion matrix ($\gls*{errRate}$, $\gls*{f05score}$, \ac{MCC}). Additionally, a study of the effect of varying the values of the feature extraction parameters of frame length and number of filters in the filter-bank was presented in order to examine their impact on the classification performance. 
The results of our work enable the extraction, from \ac{AE1} signals, of spectrum features and \ac{CC} focused on specific frequency bands that were shown to carry significantly discriminant information. The answers to the four research questions that were defined in the introduction, as the main aim of this work, are summarised in the following two paragraphs.

The results show that reducing the dimensionality of the \ac{STFT} spectrum using a mel-spaced triangular filter-bank improves the classification performance compared to using the full resolution spectrum. Furthermore, taking the natural logarithm of the \ac{STFT} spectrum and subsequently computing the \ac{DCT} can also improve the performance. The analysis signifies that the results are also improved when using mel-frequency scaling instead of linear frequency. In fact, using \ac{CART}, \ac{LDA} and \ac{SVM} as the tools for classification, the findings demonstrate that low order coefficients from the $\gls*{mfccfeatSet}$ feature set (especially the $\boldsymbol{f}_3^M$ feature vector) can distinguish between 
healthy and \ac{OA} knees with the highest \ac{AUC} amongst the 5 feature sets examined.

The experiments conducted to investigate the effect of the frame length and the number of filters, revealed two frequency regions, namely $220$ to $420$~Hz and $1$ to $3.4$~kHz. These regions contain a collection of features 
(both static and derivative coefficients) that individually score higher classification results than the rest in their respective feature set. The analysis performed highlighted the importance for classification performance of the spectrum features within these two bands. 
Finally, the very good classification performance in the experiments (Table~\ref{table:svmlinAvgRes}, Fig.~\ref{fig:auc24stft} to Fig.~\ref{fig:nbandsRes_2}) validates the hypothesis outlined at the beginning of Section~\ref{sec:featExtSection} which stated that the acoustic artifacts caused by walking on the treadmill are uncorrelated with the analysed features.

Several fruitful directions for future research can be identified from the conclusions described above. For example, the combination of features from different domains was not explored for classification as the research interest was focused at investigating the feature domains independently in order to identify which ones contained more descriptive information of \ac{OA}. Therefore, feature domain combination can be an interesting topic for future research. In addition, given the results and the insights derived from the experiments it follows that it will be interesting to examine whether an adaptive filter-bank can be beneficial for the analysis.

Contrary to other studies that focus on sit-to-stand movements and similar variants (e.g. knee flexion and extension), this study analysed signals obtained from knees performing a dynamic action. The outcomes presented in this paper suggest that the analysis of such signals can lead to non-invasive detection of knee \ac{OA} with high accuracy and could potentially extend the range of available tools for the assessment of the disease as a more practical and cost effective method without requiring clinical setups.


%

\section*{Acknowledgment}
The authors would like to thank the scientists and engineers at the MSk Lab in Charing Cross Hospital (Department of Surgery and Cancer, Faculty of Medicine, Imperial College London) for setting up the data acquisition system, recruiting the patients and acquiring the sound signals used in this study.

\ifCLASSOPTIONcaptionsoff
  \newpage
\fi

\IEEEtriggeratref{50}


\bibliographystyle{IEEEtran}
\bibliography{IEEEabrv,./sapref}
\end{document}